\documentclass[11pt,a4paper]{article}
\pdfoutput=1 
\usepackage{jheppub_mod}
\usepackage[utf8]{inputenc}
\usepackage[T1]{fontenc}
\usepackage{amsmath,amsfonts,amssymb,physics,subcaption}
\usepackage{mathtools}
\usepackage[usenames,dvipsnames]{xcolor}
\usepackage{setspace}
\usepackage{verbatim}
\usepackage{graphicx}
\usepackage{booktabs}
\usepackage{stix}
\usepackage{bm}
\usepackage{tikz}
\usepackage{amsthm}
\newtheorem{proposition}{Proposition}
\newtheorem{corollary}{Corollary}
\newlength{\PanelHeight}
\newcommand{\ds}{$dS_3\times \mathbb{R}$ }

\newcommand{\mck}{\hat{\mathcal{K}}^{(\kappa)}}
\newcommand{\jck}{\hat{\mathcal{J}}^{(\kappa)}}
\newcommand{\pp}{{\hat{p}}}
\def\l{\left}
\def\r{\right}

\title{Maximally Symmetric Boost-Invariant Solutions of the Boltzmann Equation in Foliated Geometries}
\author[a]{Mauricio Martinez,}
\author[b]{Christopher Plumberg}
\affiliation[a]{Department of Chemistry and Biochemistry, 103 Chemistry and Biochemistry Building, Bozeman, 59717, MT, USA}
\affiliation[b]{Natural Science Division, Pepperdine University, Malibu, CA 90263, USA}

\date{October 2025}
\abstract{In this work we study the relativistic kinetic theory of a boost-invariant conformal gas on a static, maximally symmetric background \ds, considering all constant-curvature slicings of $dS_3$ - flat, spherical, or hyperbolic- and their associated symmetry groups.
Using a symmetry-driven cotangent-bundle approach, we show that the isometry group of each slicing acts on phase space in such a way that only its Casimir invariants and the time-like coordinate unconstrained, so the distribution function depends solely on these quantities. This yields a unified boost-invariant exact solution of the Boltzmann equation valid for each constant--curvature foliation of \ds. Specializing this general solution to the flat and spherical foliations reproduces the Bjorken and Gubser flows, respectively, while its restriction to the hyperbolic foliation produces a genuinely new analytic solution (`Grozdanov flow'). Hydrodynamics and free streaming emerge naturally as limiting regimes of this novel exact solution. We further comment on several relevant aspects of the new boost-invariant solution on the hyperbolic slicing and on their interpretation once mapped back to Minkowski space.
}
\begin{document}

\maketitle

\section{Introduction}
\label{sec:intro}
The study of linear and nonlinear equations of motion lies at the foundation of modern theoretical and computational physics. Such equations provide the essential framework for describing, modeling, and simulating complex dynamical phenomena across a wide range of scales. In particular, nonlinear transport processes are commonly approached through kinetic theory, where the Boltzmann equation plays a central role as the microscopic description underlying the emergent macroscopic evolution.  

The Boltzmann equation is an inherently high-dimensional integro-PDE governing the seven-dimensional on-shell one-particle distribution function, and its solutions typically require computationally intensive numerical methods. Exact analytical solutions are exceptionally rare, yet they provide powerful insight into nonlinear and far-from-equilibrium physics. The limited number of known solutions, both in non-relativistic~\cite{KW-1,KW-2,bobylev} and relativistic~\cite{Denicol:2014xca,Denicol:2014tha,Bazow:2015dha,Bazow:2016oky,Wang:2025wyh,Hatta:2015kia,Noronha:2015jia,Hu:2024utr,Chen:2023vrk,Baym:1984np} settings, arise only under special circumstances, such as high degrees of symmetry or simplified collision kernels. Beyond their theoretical importance, exact solutions of the Boltzmann equation provide indispensable benchmarks for testing, validating, and calibrating sophisticated Boltzmann transport algorithms, see Ref.~\cite{Tindall:2016try}. 

Recently, Grozdanov introduced an elegant construction of both inviscid and viscous boost-invariant hydrodynamic solutions arising from a \textit{hyperbolic slicing} of \ds, hereafter referred to as the `Grozdanov flow'. Building on the geometric method of Gubser and Yarom~\cite{Gubser:2010ui}, his analysis clarifies that the Bjorken~\cite{Bjorken:1982qr} and Gubser~\cite{Gubser:2010ze} flows are not related by a generalisation of the former, but instead belong - together with his new boost invariant solution - to a unified family of three maximally symmetric boost-invariant conformal flows. These correspond respectively to the flat $ISO(2)$, spherical $SO(3)$ and hyperbolic $SO(2,1)$ foliations of the same auxiliary \ds geometry~\footnote{In Gubser’s original formulation~\cite{Gubser:2010ze}, the symmetry group is denoted as $SO_q(3)$ where $q$ is an arbitrary energy scale. Here we omit the subscript $q$, while keeping in mind that this notation refers to the same underlying flow solution.}. From this novel viewpoint, the Grozdanov flow is on equal conceptual footing with Bjorken and Gubser: all three are coordinate projections of a single geometric construction, each preserving a three-dimensional non-Abelian isometry group.

Motivated by the geometric insights underlying the Grozdanov flow, we develop in this work the kinetic theory of an uncharged conformal relativistic gas evolving on a maximally symmetric \ds background. Adopting Grozdanov’s unifying perspective, we formulate the microscopic description of the Bjorken~\cite{Bjorken:1982qr}, Gubser~\cite{Gubser:2010ze}, and Grozdanov~\cite{Grozdanov:2025cfx} flows within a single kinetic–theory framework. Our results follow from applying the cotangent–bundle formulation of relativistic kinetic theory~\cite{Rioseco:2016jwc,rioseco2019relativistic,Acuna-Cardenas:2021nkj,Sarbach:2013fya,Sarbach:2013hna,Sarbach:2013uba}, which makes transparent the role of symmetry generators and Casimir invariants in determining the structure and dynamics of an expanding gas on any foliation of \ds and in reducing the complexity of the Boltzmann equation. Solving the Boltzmann equation directly in this unified geometry allows us to recover the known kinetic solutions for Bjorken~\cite{Baym:1984np} and Gubser flows~\cite{Denicol:2014xca,Denicol:2014tha} and, crucially, to obtain a new exact solution describing a gas undergoing Grozdanov flow. This unified viewpoint also clarifies how distinct macroscopic behaviors—hydrodynamical and non-hydrodynamic free streaming regimes—emerge from the same microscopic description in different dynamical limits.

We have organized the paper as follows. 
Section~\ref{sec:cotangent} introduces the cotangent-bundle formulation of kinetic theory and explains how symmetry choices constrain the distribution function. Section~\ref{sec:exactsols} applies this framework to the Boltzmann equation on foliated \ds geometries, deriving the corresponding symmetry generators, Casimirs, reduced kinetic equation, and exact solutions for each foliation. Section~\ref{sec:exactmacro} analyzes the macroscopic dynamics emerging from these solutions, including ideal, viscous, and second-order hydrodynamics, as well as free streaming. Section~\ref{sec:dstoMink} highlights key features of the \ds solutions when mapped back to Minkowski space. Conclusions are presented in Section~\ref{sec:concl}, with technical details provided in the appendices.

Before proceeding, we clarify our notation. A hat (the symbol `\textasciicircum') written above a variable or tensor denotes that the corresponding quantity is defined in \ds or $dS_3$ space. To distinguish scalars, tensors, and other entites corresponding to different foliations of $dS_3$, we use the label $\kappa$ (or $(\kappa)$) as a subscript or superscript~\footnote{The choice is purely notational and context-dependent. For example, writing $\hat{A}_i^{(\kappa)}$ emphasizes that $\hat{A}_i$ is a vector associated with the $\kappa$ foliation (rather than a component of a second-rank tensor), whereas using $d_\kappa$ (or $d^{(\kappa)}$) simply indicates that the scalar $d$ on that particular foliation.}. The metric signature follows the `mostly plus' convention. 

\section{Cotangent bundle formulation of kinetic theory}
\label{sec:cotangent}
 In this section, we briefly review the cotangent bundle formulation of relativistic kinetic theory on a generic curved manifold. Beyond its elegant geometric structure, this formulation offers a transparent way to understand how symmetries constrain the functional form of the distribution function. A comprehensive discussion of this formalism can be found in Refs.~\cite{Rioseco:2016jwc,rioseco2019relativistic,Acuna-Cardenas:2021nkj}. An alternative, albeit equivalent, description formulated for the tangent bundle can be found in Refs.~\cite{Sarbach:2013fya,Sarbach:2013hna,Sarbach:2013uba} and references therein. The reader primarily interested in the symmetry constraints of the distribution function in \ds may proceed directly to Sect.~\ref{subsec:canlift}.

\subsection{Basics of relativistic kinetic theory}
\label{subsec:basics}

Relativistic kinetic theory~\cite{DeGroot:1980dk,Cercignani} in a curved spacetime can be naturally formulated on the cotangent bundle
\(\mathcal{T}^*\mathcal{M}\) of the spacetime manifold \(\mathcal{M}\)~\cite{Rioseco:2016jwc,rioseco2019relativistic,debbasch2009general1,debbasch2009general2}. Within this formulation, each point of \(\mathcal{T}^*\mathcal{M}\) represents a spacetime event \(x^\mu\in \mathcal{M}\) together with a co-vector 
\(p_\mu\) at this event. $p_\mu$ represents the canonical momentum of the particle at $x^\mu$.\footnote{
The zeroth component of the momentum is determined via the on-shell massless condition, i.e., $p_0=p_0(g_{\mu\nu},x^\mu,p_i)$.} The one-particle distribution function \(f(x^\mu,p_i)\) is thus defined as a non-negative function of the seven-dimensional on-shell mass submanifold of \(\mathcal{T}^*\mathcal{M}\),
\begin{equation}
\Gamma =\{(x^\mu,p_i): x^\mu\in \mathcal{M}, p_i\in \mathcal{T}^*_x \mathcal{M}|\, g_{\mu\nu}p^\mu p^\nu=0,\; p^0>0\};
\label{eq:gspace}
\end{equation}
In other words, $\Gamma$ is the massless on-shell submanifold of the cotangent bundle $\l(\mathcal{T}^*\mathcal{M}\r)$ with positive timelike momentum component $p^0$ and thus constitutes the full 7-dimensional relativistic phase space. Throughout we assume that the spacetime $(\mathcal M,g)$ is \emph{time-oriented}. Equivalently, there exists a smooth nowhere-vanishing timelike vector field $T^\mu$ that selects a
consistent choice of future lightcone at every point.
The condition $p^0>0$ in \eqref{eq:gspace} is then understood as the coordinate expression of the
\emph{future-directed} branch of the mass shell: in invariant terms one restricts to covectors $p\in T_x^*\mathcal M$
satisfying $g^{\mu\nu}p_\mu p_\nu=0$ and $p(T)<0$ (mostly-plus convention), which in a time-adapted chart
reduces to $p^0>0$.
In addition, the cotangent bundle $\mathcal T^*\mathcal M$ carries the canonical (Liouville) one-form
\begin{equation}
\vartheta \;=\; p_\mu\,dx^\mu ,
\end{equation}
and the associated canonical symplectic form
\begin{equation}
\omega \;:=\; d\vartheta \;=\; dx^\mu\wedge dp_\mu .
\label{eq:canon_sympl}
\end{equation}
This two-form is intrinsic (coordinate-independent): it is preserved by the cotangent lift of any
diffeomorphism of $\mathcal M$, and in particular by the lifted action of any spacetime isometry group.
Moreover, for any Hamiltonian $\mathcal{H}$ on $\mathcal T^*\mathcal M$, the Hamiltonian flow generated by $\mathcal{H}$
preserves $\omega$ (equivalently $\pounds_{X_H}\omega=0$).
Since the on-shell manifold $\Gamma\subset \mathcal T^*\mathcal M$ is seven-dimensional, it is not itself
symplectic; instead it inherits the pullback $i^*\omega$ (with $i:\Gamma\hookrightarrow \mathcal T^*\mathcal M$),
which is a closed but generally degenerate two-form (a presymplectic structure).
Accordingly, we regard $\omega$ on $\mathcal T^*\mathcal M$ as the fundamental symplectic structure and
restrict the dynamics and symmetry constraints to $\Gamma$ when imposing the mass-shell condition.

In this picture, the evolution of the system is governed by the
Liouville vector field \(\mathcal{L}\),
\begin{equation}
\mathcal{L} = p^\mu \frac{\partial}{\partial x^\mu}
    + \Gamma _{\mu i}^{\lambda }p_{\lambda }p^{\mu} \frac{\partial}{\partial p_i}\,,
    \label{eq:liouville}
\end{equation}
which generates the free--streaming flow in phase space and is tangent to \(\mathcal{T}^*\mathcal{M}\). In the presence of collisions among particle constituents, the relativistic Boltzmann equation can be written compactly as
\begin{equation}
    \mathcal{L}[f] = \mathcal{C}[f]\,,\quad
\Rightarrow\,\,p^{\mu }\partial _{\mu }f+\Gamma _{\mu i}^{\lambda }p_{\lambda }p^{\mu}
\frac{\partial f}{\partial p_{i}}=\mathcal{C}[f]\,,
\label{eq:Boltzmann}
\end{equation}
where $\mathcal{C}[f]$ is the collisional kernel. In this work we model the collisional kernel $\mathcal{C}[f]$ to be given by the relaxation time approximation (RTA)~\cite{BGK,Anderson:1974}
\begin{equation}
    \label{eq:rta}
    \mathcal{C}[f] =\frac{p\cdot u}{\tau_{r}}\,\left[f(x^\mu,p_i)-f_{eq.}(x^\mu,p_i,T)\right]\,.
\end{equation}
In the previous expression $T$ is the temperature in the local rest frame, $u^\mu$ is the fluid velocity, $\tau_r$ is the relaxation time scale and $f_{eq.}$ is the J\"uttner thermal equilibrium distribution function. The fluid velocity is chosen to be defined in the Landau frame. In addition, Weyl invariance demands that the relaxation time be inversely proportional to the temperature so $\tau_r =\gamma/T$. For the relaxation time approximation~\eqref{eq:rta} the proportionality constant is related to the shear viscosity over entropy ratio as follows 
\begin{equation}
    \gamma = 5\,\frac{\eta}{\mathcal{S}} \Rightarrow \tau_r = 5\frac{\eta}{\mathcal{S}}\,\frac{1}{T}\,.
    \label{eq:RTArel}
\end{equation}
Although the previous derivation is obtained from first principles~\cite{Denicol:2011fa}, in numerical studies of the RTA Boltzmann equation~\cite{Florkowski:2013lya,Florkowski:2013lza,Denicol:2014tha,Denicol:2014xca}, it is common to regard $\eta/S$ as a tunable parameter, enabling one to explore the crossover from nearly ideal to strongly viscous dynamics. In this setting, the relation~\eqref{eq:RTArel} is often rewritten as $\tau_r=c\,\eta/\left(\mathcal{S}T\right)$ with $\eta/\mathcal{S}$ taken as a fixed input and $c$ regarded as an adjustable constant. In this work we follow this convention. For a given solution of the Boltzmann equation, it is possible to determine macroscopic physical observables by considering the moments of the distribution function. For instance, the energy momentum tensor reads as
\begin{equation}
    \label{eq:tmn}
    \begin{split}
         T^{\mu\nu}=\langle\,p^\mu\,p^\nu\,\rangle\,,
    \end{split}
\end{equation}
where we denote the momentum moment of an arbitrary physical observable $\mathcal{O}(x^\mu,p_i)$ as \\$\langle \mathcal{O}(x^\mu,p_i)\rangle_X =\int_{p}\mathcal{O}(x^\mu,p_i) f_X(x^\mu,p_i)$ being the momentum measure $\int_p\equiv\int d^3p/\left[(2\pi)^3\,\sqrt{-g}\,p^0\right]$ being $\sqrt{-g}$ the square root of the determinant of the metric. The energy density $\varepsilon$, pressure $\mathcal{P}$ and shear-stress tensor $\pi^{\mu\nu}$ are read from the previous expression by splitting the momentum into its spatial and temporal components, i.e., $p^\mu = -(u\cdot p)u^\mu +p^{\langle \mu\rangle}$ with $p^{\langle \mu\rangle}=\Delta^{\mu\nu}p_\nu$ where $\Delta^{\mu\nu} = g^{\mu\nu}+u^\mu u^\nu$ is a projector orthogonal to the fluid velocity $u^\mu$. By using this decomposition of the particle's momentum, the energy-momentum tensor~\eqref{eq:tmn} is written as
\begin{equation}
    \label{eq:tmn-decomp}
    T^{\mu\nu} = \epsilon(x^\mu)u^\mu u^\nu +\Delta^{\mu\nu}(x^\mu)\mathcal{P}(x^\mu)+\pi^{\mu\nu}(x^\mu)\,,
\end{equation}
where
\begin{subequations}
\label{eq:tmn-components}
\begin{align}
\label{eq:energy}
\varepsilon (x^\mu)& =\langle\,(-u\cdot p)^{2}\,\rangle\,,   
\\
\label{eq:press}
\mathcal{P}(x^\mu)& =\frac{1}{3}\langle\,
\Delta _{\mu\nu }p^{\nu }p^{\mu }\,\rangle\,,\\
\pi ^{\mu \nu }(x^\mu)& =\langle\,
p^{\langle \mu}p^{\nu \rangle }\,\rangle\,.\label{eq:shear}
\end{align}
\end{subequations}
In Eq.~\eqref{eq:shear} we use the shorthand notation $p^{\langle \mu }p^{\nu
\rangle }{\,=\,}\Delta _{\alpha \beta }^{\mu \nu }p^{\alpha }p^{\beta }$, where the projector 
\begin{equation*}
    \Delta _{\alpha \beta }^{\mu \nu }=\frac{1}{2}\,\left( \Delta _{\alpha }^{\mu }\Delta
_{\beta }^{\nu }+\Delta _{\beta }^{\mu }\Delta _{\alpha }^{\nu }-\frac{2}{3}
\Delta ^{\mu \nu }\Delta _{\alpha \beta }\right) \,,
\end{equation*}
extracts the traceless, $u^\mu$-orthogonal component of a rank-2 tensor.
\subsection{Symmetry Constraints on the Distribution Function}
\label{sec:sym}
A spacetime symmetry generated by a Killing vector field \(\mathcal{K}\) of \(\mathcal{M}\)
induces a canonical lift to the cotangent bundle \(\mathcal{T}^*\mathcal{M}\), that is, a natural extension of \(\mathcal{K}\) from spacetime to phase space, which acts not only on the spacetime coordinates but also on the momentum components in a way
that preserves the canonical symplectic structure. In adapted local coordinates,\footnote{
By \emph{adapted local coordinates} we refer to the coordinates $(x^\mu,p_\mu)$ which provide at each point $(x,p) \in \mathcal{T}^*\mathcal{M}$ a basis of vector fields $\{\partial_{x^\mu}|_{(x,p)},\partial_{p_\mu}|_{(x,p)}\}$ and associated dual basis of covector fields $\{dx^\mu_{(x,p)},(dp_\mu)_{(x,p)}\}$~\cite{Acuna-Cardenas:2021nkj}. Hence, any vector field $V$ in $\mathcal{T}^*\mathcal{M}$ can be expanded locally as
\begin{equation*}
    V = X^\mu \frac{\partial}{\partial x^\mu}+Y_\mu\frac{\partial}{\partial p_\mu}\,,\qquad X^\mu = dx^\mu(V)\,,\qquad Y_\mu = dp_\mu(V).
\end{equation*} } the canonical lift of a spacetime Killing vector field 
$\mathcal{K}$ to the cotangent bundle takes the form~\cite{Rioseco:2016jwc,rioseco2019relativistic,Acuna-Cardenas:2021nkj}
\begin{equation}
\label{eq:lift}
\begin{split}
\tilde{\mathcal{K}} &= \mathcal{K}^\mu \frac{\partial}{\partial x^\mu}
  - p_\alpha\frac{\partial\mathcal{K}^\alpha}{\partial x^\mu}\frac{\partial}{\partial p_\mu}\,,\\
\end{split}
\end{equation}
which acts on the distribution function as a directional derivative in phase space. It is possible to rewrite the previous expression as follows
\begin{equation}
    \label{eq:liftgen}
    \tilde{\mathcal{K}} 
  = \mathcal{K}^\mu \frac{D}{\partial x^\mu}
  - p_\alpha (\nabla_\nu \mathcal{K}^\alpha)
    \frac{\partial}{\partial p_\nu}\,.
\end{equation}
The operators appearing in~\eqref{eq:liftgen} decompose the geometry of the cotangent 
bundle into \emph{horizontal} and \emph{vertical} directions. The horizontal basis vector
\begin{equation}
    \frac{D}{\partial x^\mu} 
    = \frac{\partial}{\partial x^\mu}
    + \Gamma^\alpha_{\mu\nu} p_\alpha 
      \frac{\partial}{\partial p_\nu},
\end{equation}
represents infinitesimal displacements in spacetime that are \emph{lifted} to phase space so that momenta are parallel transported along the underlying connection. In contrast, vertical basis vectors $\partial / \partial p_\mu$ generate pure 
variations in momentum at fixed spacetime position. 
Together, $\{D/\partial x^\mu,\, \partial/\partial p_\nu\}$ provide a covariant 
splitting of the tangent space of the cotangent bundle, 
$T(T^*M) = H(T^*M) \oplus V(T^*M)$, into horizontal (spacetime) and vertical (momentum) directions. Finally, the covariant derivative of the Killing vector field $\mathcal{K}^\lambda$ in Eq.~\eqref{eq:liftgen} is given by
\begin{equation}
    \nabla_\nu \mathcal{K}^\lambda 
    = \partial_\nu \mathcal{K}^\lambda 
    + \Gamma^\lambda_{\nu\sigma}\mathcal{K}^\sigma,
\end{equation}
so that the second term in the RHS of Eq.~\eqref{eq:liftgen} accounts for the induced change in momentum 
caused by the variation of $\mathcal{K}$ along the connection. 
This construction ensures that $\widehat{\mathcal{K}}$ acts intrinsically on 
phase-space functions without reference to any coordinate-dependent artifact.
Now, the associated momentum map,
\begin{equation}
\mathcal{J}_\mathcal{K} = p_\mu \mathcal{K}^\mu,
\label{eq:mom-map}
\end{equation}
remains constant along particle trajectories. In other words,
$\mathcal{J}_\mathcal{K}$ is conserved along the geodesic Liouville flow
when $\mathcal{K}$ is a Killing vector field, since the Liouville operator
acts on $\mathcal{J}_\mathcal{K}$ as $\mathcal{L}[\mathcal{J}_\mathcal{K}] = 0$.
A distribution function is said to \emph{share the symmetry} generated by \(\mathcal{K}\)
when it is invariant under the lifted flow, that is,
\begin{equation}
\pounds_{\tilde{\mathcal{K}}} f = 0.
\label{eq:LieK}
\end{equation}
where  $\pounds_X$ refers to the Lie derivative with respect to the vector field $X$. Equation~\eqref{eq:LieK} expresses that \(f\)
is constant along the integral curves of \(\widehat{\mathcal{K}}\)
and depends only on the invariants of the symmetry group. In practice, the condition of invariance of the distribution function~\eqref{eq:LieK} along the flow generated by a Killing vector reduces to a first-order linear PDE~\footnote{The distribution function $f=f(x^\mu,p_i)$ is defined in the mass shell section of the total phase space $\gamma_m$~\eqref{eq:gspace}, so $p_0$ is constrained by the on-mass shell condition. The lifted Killing vector $\widehat{\mathcal{K}}$ is tangent to $\gamma_m$ since $\widehat{\mathcal{K}}(g_{\mu\nu}p^\mu p^\nu)=0$ for any Killing vector field. Hence, the invariance condition~\eqref{eq:LieK} remains valid when $f$ is restricted to be on-shell variables.}
\begin{equation}
    \label{eq:liftPDE}
    \mathcal{K}^\mu \frac{\partial f}{\partial x^\mu} -p_\alpha\,\frac{\partial K^\alpha}{\partial x^\mu}\frac{\partial f}{\partial p_\mu}=0\,.
\end{equation}
This equation can be interpreted as the vanishing of the total derivative of $f$ along the flow generated by $\widehat{\mathcal{K}}$ in phase space. By introducing an affine parameter $\lambda$ along integral curves of this vector field, we impose 
\begin{equation}
    \label{eq:totalder}
    \frac{df}{d\lambda}=\frac{dx^\mu}{d\lambda}\frac{\partial f}{\partial x^\mu}+\frac{dp_\mu}{d\lambda}\frac{\partial f}{\partial p_\mu} = 0\,, 
\end{equation}
where we identify the characteristic equations 
\begin{subequations}
    \label{eq:characteristic}
    \begin{align}
    \label{eq:characx}
        \frac{dx^\mu}{d\lambda}&= \mathcal{K}^\mu\,,\\
        \label{eq:characp}
        \frac{dp_\mu}{d\lambda}&=-p_\alpha\,\frac{\partial \mathcal{K}^\alpha}{\partial x^\mu}\,.
    \end{align}
\end{subequations}
The integrals of this system yield the momentum–map invariants $\mathcal{J}_\mathcal{K}$~\eqref{eq:mom-map}, associated with the Killing vector field~$\mathcal{K}$. Now, in the case when we have a set of Killing vectors \(\mathcal{K}_a\) ($a=1,..,N$) forming a Lie algebra $\mathfrak{g}$ of a symmetry group $G$,
the corresponding conditions
\(\pounds_{\tilde{K}_a} f=0\)
define a system of first–order characteristic equations in phase space. In the adapted local coordinates $(x^\mu,p_\nu)$ on $\mathcal{T}^*\mathcal{M}$, the
invariants obtained from the characteristic system of the lifted
Killing fields coincide with the momentum maps associated with each
symmetry generator, $\mathcal{J}_{a} = p_\mu \mathcal{K}^\mu_a$. Under the Poisson bracket, the momentum maps satisfy the same Lie–algebra relations as the Killing vectors from which they are constructed. Now, by differentiating $\mathcal{J}_{a}$ along the lifted flow, i.e., along a given trajectory, yields
\begin{equation}
    \frac{d\mathcal{J}_{a}}{d\lambda} = p_\mu \frac{d\mathcal{K}^\mu_a}{d\lambda}+\mathcal{K}^\mu_a\,\frac{dp_\mu}{d\lambda}\,= 0\,,
\end{equation}
where we make use of Eq.~\eqref{eq:characp}. This expression shows that the quantities derived through the characteristic equations
in adapted local coordinates are precisely these momentum maps expressed
in that basis. Hence, the condition
$\pounds_{\tilde{\mathcal{K}}_a}f=0$ admits a solution which can be written as a function of the corresponding momentum maps,
\begin{equation}
f(x^\mu,p_i) = f\big(\mathcal{J}_{1}(x^\mu,p_i),\,\mathcal{J}_{2}(x^\mu,p_i),\,\ldots,\,\mathcal{J}_{N}(x^\mu,p_i)\big).
\label{eq:invariantform}
\end{equation}
This expression reflects the fact that symmetry considerations restrict the distribution function to depend only on the corresponding momentum maps.
The characteristic equations identify the phase-space directions along which the distribution function is constant, but they do not determine which combinations of variables are globally invariant under the full symmetry group.
In other words, Eq.~\eqref{eq:invariantform} indicates that the distribution function remains constant along the phase-space trajectories generated by the infinitesimal symmetries of the system. However, this local invariance does not imply full invariance under the finite action of the symmetry group. The set $\{\mathcal{J}_{a}\}$ transforms among itself under the coadjoint representation of the group~\footnote{The coadjoint representation describes how a symmetry group acts on its own conserved quantities, while the Casimirs are the combinations of those quantities that remain invariant under any such action. Geometrically, this means that while the momentum maps label individual orbits of the coadjoint action, the Casimirs are the quantities that remain constant across each orbit, defining the invariant geometry of the phase-space manifold. For example, when the rotation group $SO(3)$, acts on space, it rotates the components of the angular momentum, changing their direction but not their magnitude.That invariant magnitude—the length of the angular momentum vector—is precisely the Casimir of $SO(3)$.}, so only particular combinations of them - the Casimir invariants of the algebra - remain globally invariant under the full group action. We elaborate on this in the following subsection.

\subsubsection{Symmetry reduction on the mass shell and Casimir dependence}
\label{subsub:symm-red}

Eq.~\eqref{eq:invariantform} is a local solution for the distribution 
function invariant under the lifted action of the symmetry group~$G$ on 
phase space, written in terms of the momentum 
maps~$\mathcal{J}_a$. However, this result reflects only infinitesimal 
invariance along the orbits of the lifted Killing fields: the individual 
momentum maps~$\mathcal{J}_a$ transform among themselves under the 
coadjoint representation of~$G$, and therefore are not separately 
invariant under the full (finite) group action. Only specific 
combinations---the Casimir invariants of the Lie algebra---are globally 
preserved.

A natural question is then: under what conditions do the Casimirs 
\emph{exhaust} all independent invariants, so that the distribution 
function~$f$ depends on nothing else? In general, the orbit quotient 
space $\Gamma/G$ can have positive 
dimension,\footnote{In App.~\ref{app:MW-review} we review the Marsden--Weinstein reduction technique as well as the definitions of cohomogeneity.} 
admitting additional $G$-invariant functions beyond the Casimirs---for 
instance, angle variables on Liouville tori in integrable systems that are not removed by the spacetime isometries, cf.~\cite{Rioseco:2016jwc}. The additional geometric hypothesis that eliminates this residual freedom is that the lifted $G$-action has cohomogeneity zero: after fixing the values of the Casimirs, $G$ acts transitively on each connected component of the 
resulting level set in phase space, so the quotient is discrete and no 
further continuous invariants 
survive~\cite{berndt2017cohomogeneity,diaz2017cohomogeneity}. We 
formalize this as follows.

\begin{proposition}[\textit{Symmetry--determined functional dependence}]
\label{prop:cohom0}
Let $(\Gamma,\omega)$ be the on--shell phase space (mass shell) with 
Hamiltonian $\mathcal{H}$ generating the Liouville flow, and let a Lie 
group $G$ act on $\Gamma$ by symplectomorphisms with equivariant 
momentum map $J:\Gamma\to\mathfrak{g}^*$, such that 
$\{\mathcal{H},\,\mathcal{J}_a\}=0$ for all~$a$. Let 
$\mathcal{N}\subset\Gamma$ be a $G$--invariant submanifold preserved by 
the Liouville flow. Suppose that, after fixing the coadjoint invariants 
(Casimirs) $\{\mathcal{C}_i\}$ built from the momentum 
maps~$\mathcal{J}_a$, the lifted $G$--action has cohomogeneity zero on 
each connected component of~$\mathcal{N}$.

Then any smooth distribution $f:\mathcal{N}\to\mathbb{R}$ satisfying 
$\pounds_{\tilde{\mathcal{K}}_a}\,f=0$ for all~$a$ can be written 
locally as
\begin{equation}
  f(x^\mu,p_i) 
  = F\!\big(\mathcal{C}_{1}(\{\mathcal{J}_{a}\}),\,
    \mathcal{C}_{2}(\{\mathcal{J}_{a}\}),\,\ldots\big)\,,
\label{eq:invariantsol}
\end{equation}
where $F$ is an arbitrary smooth function.
\end{proposition}

The proof, based on the Marsden--Weinstein symplectic reduction 
procedure~\cite{MarsdenWeinstein1974,MarsdenRatiuBook,AbrahamMarsden,OrtegaRatiu,Meyer1973,cushman2015global} 
combined with the cohomogeneity-zero 
condition~\cite{berndt2017cohomogeneity,diaz2017cohomogeneity}, is given 
in App.~\ref{app:proof-proposition}~\footnote{We briefly review the Marsden--Weinstein symplectic reduction method and cohomogeneity concept in App.~\ref{app:MW-review}.}. The cohomogeneity-zero condition is 
automatically satisfied when the isometry group~$G$ acts transitively on 
the spatial submanifold of~$\mathcal{M}$. In that case, the cotangent 
lift of~$G$ sweeps out orbits whose codimension in $T^*\mathcal{M}$ 
equals the number of independent Casimirs, and no further continuous 
invariants survive once the Casimir values are fixed. In applications, 
the only nontrivial step is to verify the stated transitivity for the 
specific lifted action on the relevant invariant subset. For the 
foliations of \ds studied in this work, the groups 
$H_\kappa\in\{SO(3),\,ISO(2),\,SO(2,1)\}$ act transitively on the 
two-dimensional constant-curvature slices~$\Sigma_\kappa$, ensuring that 
this condition holds, as shown in Corollary~\ref{cor:reduced-form} 
(proven in App.~\ref{app:proof-corollary}).

\subsection{Collisional kernel and symmetry constraints}
\label{subsec:coll}
A natural question within the cotangent bundle formalism is whether this symmetry reduction extends to the collision term, and to what extent the symmetries of the distribution function constrain the functional structure of the collisional kernel. The answer is now well understood~\cite{Acuna-Cardenas:2021nkj}: invariance of the distribution function alone, Eqs.~\eqref{eq:LieK} and~\eqref{eq:invariantform}, does not restrict the collision kernel. Rather, the decisive requirement is the invariance of the microscopic transition rate under the complete lift of the same symmetry group. If the transition probability $\mathcal{W}(x^\mu,p\to p')$ entering in the collisional kernel satisfies 
\begin{equation}
    \label{eq:trans-prob}
    \pounds_{\tilde{\mathcal{K}}_a}\mathcal{W}(x^\mu,p\to p') =0\,\quad\forall\,\,\tilde{\mathcal{K}}_a\,,
\end{equation}
then the collision operator $\mathcal{C}[f]$ is form-invariant under the lifted symmetry and maps symmetric distributions to symmetric distributions. This result was established rigorously by Acuña et al.~\cite{Acuna-Cardenas:2021nkj} (see Appendix~F therein): \textit{symmetry constraints on the collision term arise only from the lifted invariance of the transition rate, not from imposing $\pounds_{\tilde{\mathcal{K}}_a} f=0$ by itself}. Condition \eqref{eq:trans-prob} is automatically satisfied whenever the microscopic transition probability
$\mathcal W(x^\mu,p\to p')$ depends only on scalars constructed from the momenta and the spacetime metric at $x$, i.e.\ on invariants under the lifted isometry action. This is the case for the standard relativistic Boltzmann collision term describing elastic binary collisions, where the differential cross section (and hence the transition rate) can be written as a function of Lorentz invariants such as the Mandelstam variables $s,t,u$ built from the incoming and outgoing four-momenta. In such situations, the lifted generators act trivially on $\mathcal W$ and Eq.~\eqref{eq:trans-prob} follows~\footnote{We thank Prof. Oliver Sarbach for pointing out this issue.}. For the RTA collision kernel~\eqref{eq:rta} used in this work, this model is automatically compatible with the isometry generators associated to each \ds slicing, as shown in Sect.~\ref{subsec:constr}. 

We conclude by noting that when both symmetry requirements are satisfied---namely,
(i) invariance of the distribution function under the lifted generators,
Eqs.~\eqref{eq:LieK}, and (ii) equivariance of the collision kernel (equivalently,
invariance of the transition amplitude), Eq.~\eqref{eq:trans-prob}---the full Boltzmann
equation preserves the $G$-invariant sector and consistently reduces to a
symmetry-protected evolution equation for $f$.
If, in addition, the lifted $G$-action on the relevant on-shell invariant set is
\emph{sufficiently large} in the precise sense that, after fixing the coadjoint invariants
(Casimirs) of the momentum map $\{\mathcal J_a\}$, it acts \emph{transitively} on each
connected component (equivalently: the corresponding quotient is discrete, i.e. the action
has cohomogeneity zero on those Casimir level sets), then no further continuous invariants
remain.
In this case, the physically relevant symmetric solutions depend only on the Casimir
invariants of the symmetry group,
\begin{equation}
f(x^\mu,p_i) = f\big(\mathcal{C}_{1}(\{\mathcal{J}_{a}\}),\,\mathcal{C}_{2}(\{\mathcal{J}_{a}\}),\,\ldots\big)\,,
\label{eq:invariantsol}
\end{equation}
which guarantees full invariance under the lifted action of the symmetry group.

\section{Boltzmann equation and its general solution on foliated $dS_3\times\mathbb{R}$}
\label{sec:exactsols}

In this section, we apply the cotangent bundle formalism to derive an exact solution of the Boltzmann equation corresponding to a chosen foliation of \ds. We start by reviewing the key features of \ds and its possible foliations.

\subsection{From Minkowski $\mathbb{R}^{3,1}$ to foliated \ds}
\label{subsec:Gubser-Yarom}
The line element in Minkowski space coordinates is
\begin{align}
    ds^2 = -dt^2 + dx^2 + dy^2 + dz^2
\end{align}
For systems undergoing boost-invariant evolution, a more natural choice is the Milne coordinates
\begin{align}
  \tau = \sqrt{t^2-z^2}\,, \quad \varsigma = \frac{1}{2}\ln\l| \frac{t+z}{t-z} \r|\,,
\end{align}
for which
\begin{align}
  ds^2 = -d\tau^2 + dx^2 + dy^2 + \tau^2 d\varsigma^2 = -d\tau^2 + dr^2 + r d\phi^2 + \tau^2 d\varsigma^2
\end{align}

To map this into \ds, we first note that de Sitter space $dS_3$ (with unit radius) can be embedded in $\mathbb{R}^{3,1}$ as the hyperboloid defined by
\begin{align}
  -\l(X^0\r)^2 + \sum_{m=1}^3 \l( X^m \r)^2 = 1\,, \label{eq:hyperboloid}
\end{align}
and may be parametrized by the coordinates
\begin{align}
  X^0 = -\frac{1-q^2(\tau^2-r^2)}{2q\tau},\quad X^i = \frac{x^i}{\tau}, \quad  X^3 = \frac{1+q^2(\tau^2 - r^2)}{2 q \tau}\,,
\end{align}
where $q>0$ is an arbitrary parameter.
The line element of $dS_3$ is then expressible as
\begin{align}
  d\hat{s}^2 = \frac{1}{\tau^2}\l( -d\tau^2 + dx^2 + dy^2  \r) \label{dS3_milne}
\end{align}

As shown in \cite{Grozdanov:2025cfx}, one can distinguish three different ways to foliate the hyperboloid which preserve the maximum degree of symmetry: foliations at constant $X^0$, constant $X^3 - X^0$, and constant $X^3$.  Each leaf of a given foliation intersects the hyperboloid along a 2d (boost-invariant) surface with constant curvature $\kappa$, corresponding to flat ($\kappa=0$), spherical ($\kappa=+1$) and hyperbolic ($\kappa=-1$) foliations.  One parameterizes a given foliation by defining a time-like coordinate $\rho$ and a space-like coordinate $\theta$, such that $\rho$ indexes the leaf of the foliation and $\theta$ parametrizes the leaf itself.

If one foliates by $X^0 \equiv \sinh \rho$ and takes the remaining coordinates to be proportional to a unit vector $X^m = \cosh \rho\, r^m$ as done in \cite{Gubser:2010ui}, one finds from Eq.~\eqref{eq:hyperboloid} that the metric \eqref{dS3_milne} becomes 
\begin{align}
  d\hat{s}^2 = -d\rho^2 + \cosh^2\rho \, \l( d\theta^2 + \sin^2\theta \,d\phi^2 \r) \label{eq:dSigmaGubserdef}
\end{align}
On the other hand, one may foliate by $X^3 = \cosh \rho$ and take the remaining coordinates to be proportional to a unit vector in a lower dimensional de Sitter space:
\begin{align}
  d\hat{s}^2 = -d\rho^2 + \sinh^2\rho \, \l( d\theta^2 + \sinh^2\theta \,d\phi^2 \r) \label{eq:dSigmaGrozdanovdef}
\end{align}
The first foliation therefore corresponds to leaves with uniform positive (spherical) curvature and the second to leaves with uniform negative (hyperbolic) curvature.
The third type of foliation by $X^3 - X^0 = \frac{1}{q\tau} \equiv \frac{1}{q} e^{-\rho}$ yields a uniform space with zero (flat) curvature with metric
\begin{align}
  d\hat{s}^2 = -d\rho^2 + e^{-2\rho} \, \l( d\theta^2 + \theta^2 \,d\phi^2 \r), \label{eq:dSigmaBjorkendef}
\end{align}
which corresponds to a spatially flat 2d universe undergoing Hubble \textit{contraction} with a scale factor given by $a(\rho) = e^{-\rho}$ and $\theta$ playing the role of a radial coordinate. 

We can rewrite Eqs.~\eqref{eq:dSigmaGubserdef}-\eqref{eq:dSigmaBjorkendef} in a more compact notation by introducing
\begin{align}
\label{eq:fol-met}
    d\Sigma^2_\kappa \equiv \begin{cases}
        d\theta^2 + \sin^2\theta \,d\phi^2 & \kappa = +1 \\
        d\theta^2 + \theta^2 \,d\phi^2 & \kappa = 0 \\
        d\theta^2 + \sinh^2\theta \,d\phi^2 & \kappa = -1
    \end{cases}
\end{align}
With this notation, we can parametrize the infinitesimal interval between two events in any foliation of \ds by
\begin{equation}
    \label{eq:dsmetric}
    \begin{split}
    d\hat{s}^2_\kappa = \hat{g}_{\mu\nu}d\hat{x}^\mu d\hat{x}^\nu &= -d\rho^2\,+\,S_\kappa^2(\rho)\,d\Sigma_\kappa^2 +d\varsigma^2\,,\\
    &=-d\rho^2\,+\,S_\kappa^2(\rho)\,\left(d\theta^2+Q_\kappa^2(\theta)\,d\phi^2\right) +d\varsigma^2\,,
    \end{split}
\end{equation}
where the subindex $\kappa$ labels the specific foliation of \ds as defined above \cite{Grozdanov:2025cfx}. Here the set of coordinates $(\rho,\theta,\phi)$ parametrize $dS_3$ while $\varsigma\in (-\infty,\infty)$ parametrizes $\mathbb{R}$, and the functions $S_\kappa(\rho)$ and $Q_\kappa(\theta)$ as well as the associated coordinate ranges for each foliation are listed in Table~\ref{table:ds3defs}. Below, we denote a single leaf in the $\kappa$ foliation by $\Sigma_\kappa\subset\,dS_3$. A concise summary of the geometrical structure of 2d surfaces with constant curvature, their Killing vector fields and their Lie algebra is provided in App.~\ref{app:diffgeom}.  
\begin{table}[h!]
\centering
\caption{Definitions of $S_\kappa(\rho)$ and $Q_\kappa(\theta)$, together with the corresponding coordinate ranges for different spatial curvatures according to Ref.~\cite{Grozdanov:2025cfx}.}
\label{table:ds3defs}
\begin{tabular}{cccccc}
\toprule
$\kappa$ & $S_\kappa(\rho)$ & $Q_\kappa(\theta)$ & $\rho$ range & $\theta$ range & $\phi$ range \\
\midrule
$0$   & $e^{-\rho}$   & $\theta$ & $-\infty < \rho < \infty$ & $0 \leq \theta < \infty$ & $0 \leq \phi < 2\pi$ \\
$+1$  & $\cosh\rho$   & $\sin\theta$ & $-\infty < \rho < \infty$ & $0 \leq \theta < 2\pi$   & $0 \leq \phi < 2\pi$ \\
$-1$  & $\sinh\rho$   & $\sinh\theta$ & $0 < \rho < \infty$       & $0 \leq \theta < \infty$ & $0 \leq \phi < 2\pi$ \\
\bottomrule
\end{tabular}
\end{table}
%========================================================================

\subsection{Isometries of \ds foliations}
\label{sub:foliation}
The full isometry group of the spacetime \(\mathcal{M}=dS_3\times\mathbb{R}\) is $SO(3,1)\times E_1$ where $SO(3,1)$ group acts on the de Sitter coordinates, while the one-dimensional euclidean group $E_1$ acts along the additional $\varsigma$. The latter can be written as $E_1=T_1\rtimes O(1)$ where $T_1$ denotes the one-dimensional translation group, and $O(1)$ is the one-dimensional orthogonal group consisting of the identity and reflection through the origin. Hence, $O(1)$ encodes both orientation-preserving and orientation-reversing isometries of the line, and is isomorphic to the discrete group $\mathbb{Z}_2$, i.e. $O(1)\cong \mathbb{Z}_2$. The de Sitter space $dS_3$ can be foliated into 2d slices that preserve the maximum amount of symmetries~\cite{Grozdanov:2025cfx}. Therefore, each leaf of a foliation defines a specific spatial geometry whose associated isometries determine the invariants under the full action of the isometry group.  

The Killing vectors of the constant curvature slice $\Sigma_\kappa$ of $dS_3$ are~\footnote{The one–dimensional space $\mathbb{R}$ admits a single continuous isometry generated by the Killing vector $\partial_\varsigma$ corresponding to translations along $\varsigma$. In one dimension, the Killing equation admits no other nontrivial smooth solutions; the only additional isometry is the discrete reflection $\varsigma\to -\varsigma$, which is not connected to the identity and therefore not generated by a Killing vector.}
\begin{subequations}
\label{eq:Killvectsds3R}
\begin{align}
\label{eq:K1}
\mck_1(\theta,\phi,\kappa)&=\cos\phi\,\partial_\theta-G_\kappa(\theta)\sin\phi\,\partial_\phi\,,\\
\label{eq:K2}
\mck _2(\theta,\phi,\kappa)&=\sin\phi\,\partial_\theta+G_\kappa(\theta)\cos\phi\,\partial_\phi,\\
\label{eq:K3}
\mck _3(\theta,\phi,\kappa)&=\partial_\phi\,,\\
\label{eq:K4}
\hat{\mathcal{K}}_\varsigma =\partial_\varsigma\,,
\end{align}
\end{subequations}
where $G_\kappa(\theta) = Q_\kappa'(\theta)/Q_\kappa(\theta)$, see App.~\ref{app:diffgeom} for details. The first three terms in this expression correspond to the Killing vectors generating the isometries in the slice $\Sigma_\kappa$, while the last term represents the Killing vector associated with the one-dimensional Euclidean subgroup $E_1$. The Lie algebra of the Killing vectors~\eqref{eq:Killvectsds3R} is 
\begin{subequations}
    \label{eq:LiealgdSR}
    \begin{align}
    [\mck_A,\mck_\varsigma]&=0\,,\\
    \label{eq:Liefolia}
    [\mck_A,\mck_B] &= f_{AB}{}^{C}\,\mck_C\,,
    \end{align}
\end{subequations}
with $A,B,C =1,2,3$ and the structure constants $\{f_{12}{}^{3},f_{23}{}^{1},f_{31}{}^{2}\}=\{-\kappa,-1,-1\}$ (see App.~\ref{app:diffgeom} for details).  

The explicit form of the three Killing vector fields $\mck_A$~\eqref{eq:K1}-\eqref{eq:K3} and their respective algebra~\eqref{eq:Liefolia} illustrates how the geometry of each foliation determines the structure of the symmetry flows:

\begin{itemize}
    \item \textbf{For $\boldsymbol{\kappa=+1}$:} 
    The functions $G_{+1}(\theta)=\cot\theta$ and $Q_{+1}(\theta)=\sin\theta$ correspond to the metric of the two-sphere $\mathbb{S}^2$. 
    The Killing vectors $\hat{\mathcal{K}}^{(+1)}_A$ generate the $\mathfrak{so}(3)$ algebra of rotations; geometrically, they represent infinitesimal rotations around three orthogonal axes.
    The associated orbits are closed curves (great circles), reflecting the compact nature of the spherical foliation.

    \item \textbf{For $\boldsymbol{\kappa=0}$:} 
    Here $G_0(\theta)=1/\theta$ and $Q_0(\theta)=\theta$, corresponding to the flat metric of the Euclidean plane $\mathbb{R}^2$.
    The Killing vectors $\hat{\mathcal{K}}^{(0)}_1$ and $\hat{\mathcal{K}}^{(0)}_2$ generate translations along orthogonal Cartesian directions, while $\hat{\mathcal{K}}^{(0)}_3$ generates planar rotations.
    Together, they form the Lie algebra $\mathfrak{iso}(2)$, representing the Euclidean group of motions in the plane.

    \item \textbf{For $\boldsymbol{\kappa=-1}$:} 
    The functions $G_{-1}(\theta)=\coth\theta$ and $Q_{-1}(\theta)=\sinh\theta$ describe the metric of the hyperbolic plane $\mathbb{H}^2$. 
    The Killing vectors $\hat{\mathcal{K}}^{(-1)}_A$ generate the Lorentzian algebra $\mathfrak{so}(2,1)$, which acts transitively on $\mathbb{H}^2$.
    In this case, $\hat{\mathcal{K}}^{(-1)}_3$ corresponds to rotations around the symmetry axis, while $\hat{\mathcal{K}}^{(-1)}_1$ and $\hat{\mathcal{K}}^{(-1)}_2$ generate hyperbolic ``boost-like'' transformations that move points along geodesics of constant negative curvature.
\end{itemize}

\subsection{Symmetry Constraints on the Distribution Function in Foliated \ds }
\label{subsec:constr}

We start this section by examining the dynamical evolution along a chosen foliation leaf $\Sigma_\kappa$ of $dS_3$.
While the full de~Sitter space admits the six-dimensional isometry group $SO(3,1)$, each foliation into two-dimensional slices with constant curvature $\kappa$ preserves only a three-dimensional subgroup,
$H_\kappa \in \{SO(3),\, ISO(2),\, SO(2,1)\}$ whose Killing vectors correspond to Eqs.\eqref{eq:K1}-\eqref{eq:K3} for a given $\kappa$. Invariance under the lifted action of $H_\kappa$ restricts the distribution function to depend solely on the associated Casimir invariants, leaving its variation along the orthogonal coordinate~$\rho$ unconstrained.
Hence, the evolution along $\rho$ captures the residual dynamics after the full $SO(3,1)$ symmetry is reduced to $H_\kappa$. This reduction reflects the geometric restriction that connects global de~Sitter equilibrium to the boost-invariant flows of Bjorken~\cite{Bjorken:1982qr}, Gubser~\cite{Gubser:2010ze,Gubser:2010ui} and Grozdanov~\cite{Grozdanov:2025cfx}: once a preferred foliation is chosen, $\rho$ becomes the genuine dynamical variable governing the evolution. Mathematically, this corresponds to a contraction of the characteristic system of first-order PDEs on the cotangent bundle $\mathcal{T}^*\mathcal{M}$: the six $SO(3,1)$ generators defining the invariant flow reduce to the three Killing vector fields of $H_\kappa$, leaving $\rho$ as the single non-trivial characteristic direction along which $f$ evolves. In this sense, selecting a foliation singles out $\rho$ as the natural dynamical parameter of the reduced geometry, capturing the physical evolution that survives the symmetry reduction. One important consequence of the fact that all isometry generators of the foliated \ds leave the time–like coordinate $\rho$ invariant is that the relaxation–time approximation automatically respects these symmetries. Since the relaxation time depends only on this invariant variable, $\tau_r\equiv \tau_r(\rho)=\gamma/\hat{T}_\kappa(\rho)$ with $\gamma$ being a constant, the RTA collision term remains form-invariant under the full isometry group of the foliation. 

For an exact or approximate boost-invariant solution of the Boltzmann equation, the distribution function is constrained by the lifted action of the symmetry generators of the chosen foliation $\Sigma_\kappa$ of $dS_3$ that leave the constant-curvature slices $\kappa=0,\pm 1$ invariant, together with the $E_1$ generator along $\mathbb{R}$. The number of independent variables on which the distribution function depends is directly derived from general group-theoretical arguments as follows. For a given Lie algebra $\mathfrak{g}$,
the number of algebraically independent Casimir invariants $N_{\mathcal{C}}$
is determined by the dimension of the algebra and the rank of the antisymmetric
structure matrix $F_{ab}$ constructed from the structure
constants~\cite{beltrametti1966number,humphreys2012introduction}, namely $N_{\mathcal{C}} = \dim(\mathfrak{g}) - \mathrm{rank}(F_{ab})\,$.
For the subalgebras associated with the constant-curvature foliations of $dS_3$,
$\mathfrak{h}_\kappa \in \{\mathfrak{so}(3),\, \mathfrak{iso}(2),\, \mathfrak{so}(2,1)\}$,
the structure matrix has generic rank two, yielding a single independent Casimir,
$N_{\hat{\mathcal{C}}^{(\kappa)}} = 1$.
Similarly, for the one-dimensional Euclidean group
$E_1 = T(1)\rtimes O(1)$, the Lie algebra
$\mathfrak{e}_1\cong\mathbb{R}$ admits one invariant,
$N_{\mathcal{C}_{\varsigma}} = 1$. Hence, an exact solution of the Boltzmann equation in \ds that is invariant under the full isometry group action on a foliation leaf $\Sigma_\kappa$ depends only on three independent variables:
the time-like coordinate $\rho$, the Casimir $\hat{\mathcal{C}}^{(\kappa)}$
associated to the isometry group of $\Sigma_\kappa$, and the Casimir
$\hat{\mathcal{C}}_{\varsigma}$ corresponding to the $E_1$ sector,
\begin{equation}
\label{eq:gensol}
f = f\!\left(\rho,\,
\hat{\mathcal{C}}^{(\kappa)}(\{\hat{\mathcal{J}}^{(\kappa)}_{A}\}),\,
\hat{\mathcal{C}}_{\varsigma}(\{\hat{\mathcal{J}}_{\varsigma}\})\right)\,.
\end{equation}
A sharper formulation of this result is given in Sect.~\ref{subsec:dist-theorem}.

\subsection{Casimirs of the foliated \ds geometry}
\label{subsec:canlift}
In the following we proceed to calculate the Casimirs based on standard group theory methods~\cite{humphreys2012introduction}. 
\subsubsection{Casimir of the 2d constant curvature foliation $\Sigma_\kappa$}
\label{subsub:casds3}
The associated momentum maps of the slice $\Sigma_\kappa$ are obtained from their definition~\eqref{eq:mom-map} and the Killing vectors~\eqref{eq:K1}-\eqref{eq:K3}. These are given by
\begin{subequations}
    \label{eq:mommaps-ds}
    \begin{align}
        \jck_1&= \hat p_\theta\,\cos\phi - \hat p_\phi\, G_\kappa\,\sin\phi \,,\\
        \jck_2&= \hat p_\theta\,\sin\phi + \hat p_\phi\, G_\kappa\,\cos\phi\,,\\
        \jck_3&= \hat p_\phi\,.
    \end{align}
\end{subequations}
The functions $\jck_A$ are conserved quantities associated with the symmetry flows generated by the Killing vectors of the slice $\Sigma_\kappa$ in $dS_3$, i.e.\ the momentum maps on the cotangent bundle $T^*\Sigma_\kappa$.  
For $\kappa = +1$, they coincide with the components of the angular momentum on the spherical slice $\mathbb{S}^2$, reflecting the rotational invariance of $SO(3)$.  
For $\kappa = 0$, they represent the conserved linear and orbital momenta on the flat plane $\mathbb{R}^2$, corresponding to translations and rotations generated by $SO(2)$.  
Finally, for $\kappa = -1$, they form the components of a Lorentz vector associated with $SO(2,1)$, describing the invariants of motion under hyperbolic ``boost-like'' transformations on the negatively curved slice $\mathbb{H}^2$.  
Thus, the set $\{\jck_A\}$ encapsulates, in a unified way, the conserved dynamical quantities for all foliations of $dS_3$.

The associated momentum maps $\jck_{A}$~\eqref{eq:mommaps-ds} satisfy an identical algebra to the corresponding Killing vectors~\eqref{eq:Liefolia} under
Poisson brackets, namely 
\begin{equation}
\label{eq:LiealgJdsfol}
\{\jck_A,\jck_B\}
= f_{AB}{}^{C}\,\jck_{C},
\qquad
f_{12}{}^{3}=-\kappa,\;
f_{23}{}^{1}=f_{31}{}^{2}=-1.
\end{equation}
The quadratic Casimir is built from an $\mathrm{ad}$-invariant~ symmetric bilinear form~\footnote{Let $\mathfrak{h}$ be a Lie algebra with generators $T_a$ and structure constants $[T_a,T_b]=f_{ab}{}^c T_c$. A symmetric bilinear form $g^{ab}$ on $\mathfrak{h}$ is called ad-invariant if it is preserved under the adjoint action, i.e.
\begin{equation*}
    g\left([X,Y],Z\right)+ g\left(Y,[X,Z]\right) = 0\,,\qquad \forall\,\,X,Y,Z \in \mathfrak{h}\,.
\end{equation*}
In components, this condition is equivalent to
\begin{equation*}
    f_{ad}{}^{c}g_{cb}+f_{bd}{}^{c}g_{ac}=0
\end{equation*}
If $g_{ab}$ is ad-invariant then the quadratic element $g^{ab}T_a T_b$ commutes with all generators of $\mathfrak{h}$ (i.e. is invariant under the adjoint action) and therefore defines a quadratic Casimir of the algebra.
} $g^{ab}$ satisfying
$f_{ad}{}^{c}g_{cb}+f_{bd}{}^{c}g_{ac}=0$.
For the above commutation relations, isotropy in the $(1,2)$-plane allows
$g_{ab}=\mathrm{diag}(A,A,B)$, and the ad-invariance condition imposes 
$\kappa B - A = 0$, hence $B=A/\kappa$.
Choosing the normalization $A=1$ gives
\[
g^{ab}=\mathrm{diag}(1,1,\kappa),
\]
and the resulting Casimir function
\begin{equation}
\label{eq:Ckappa}
\hat{\mathcal{C}}_\kappa
  = \left(\jck_{1}\right)^2 +\left (\jck_{2}\right)^2 + \kappa\,\left(\jck_{3}\right)^2.
\end{equation}
The Casimir invariant $\hat{\mathcal{C}}_\kappa$ represents the invariant squared distance of the momentum map vector $(\jck_{1},\jck_{2},\mathcal{J}_{3})$ with respect to the metric of the constant-curvature slice $\Sigma_\kappa$. For $\kappa = +1$, it reduces to the familiar expression for the angular-momentum magnitude on the sphere, $\hat{\mathcal{C}}_{+1} =\left( \hat{\mathcal{J}}^{(1)}_{1}\right)^2 + \left( \hat{\mathcal{J}}^{(1)}_{2}\right)^2+\left( \hat{\mathcal{J}}^{(1)}_{3}\right)^2$, and is invariant under rotations of $\mathfrak{so}(3)$. For $\kappa = 0$, $\hat{\mathcal{C}}_0 = \left( \hat{\mathcal{J}}^{(0)}_{1}\right)^2 + \left( \hat{\mathcal{J}}^{(0)}_{2}\right)^2$ gives the squared planar momentum invariant under the Euclidean group $\mathfrak{iso}(2)$.
For $\kappa = -1$, the sign change in the third term reflects the Lorentzian signature of the hyperbolic metric, and $\hat{\mathcal{C}}_{-1} = \left( \hat{\mathcal{J}}^{(-1)}_{1}\right)^2 + \left( \hat{\mathcal{J}}^{(-1)}_{2}\right)^2 - \left( \hat{\mathcal{J}}^{(-1)}_{3}\right)^2$ defines the invariant hyperbolic norm of the momentum vector—the analogue of the Minkowski length associated with $\mathfrak{so}(2,1)$. Thus, the Casimir unifies the invariant `momentum norms' across the three foliations, providing the conserved scalar quantity that labels the group orbits in phase space.

Substituting Eq.~\eqref{eq:mommaps-ds} into the Casimir~\eqref{eq:Ckappa} yields
\begin{equation}
\hat{\mathcal{C}}_\kappa\equiv \hat{p}_{\Sigma_\kappa^2}= \hat{p}_\theta^2 + (G_\kappa^2+\kappa)\hat{p}_\phi^2 = \hat{p}_\theta^2 + \frac{\hat{p}_\phi^2}{Q_\kappa(\theta)^2},
\label{eq:Cas-fol}
\end{equation}
with $G_\kappa(\theta)=Q_\kappa'(\theta)/Q_\kappa(\theta)$ and we used the identity $G_\kappa^2+\kappa=1/Q_\kappa^2$. Replacing the function $Q_\kappa(\theta)$ for each curvature parameter (see Table~\ref{table:ds3defs}) in the previous expression yields 
\begin{subequations}
\label{eq:mom-cas}
    \begin{align}
    \label{eq:mom-casGub}
    \hat{\mathcal{C}}_1\equiv \hat{p}^2_{\Sigma_1} &= \hat{p}_\theta^2 + \frac{\hat{p}_\phi^2}{\sin^2\theta}\,\\
    \label{eq:mom-casBj}
    \hat{\mathcal{C}}_0\equiv \pp^2_{\Sigma_0}  &= \pp_\theta^2 +\frac{\pp_\phi^2}{\theta^2}\,,\\
    \label{eq:mom-casHyp}
    \hat{\mathcal{C}}_{-1}\equiv\hat{p}^2_{\Sigma_{-1}} &= \hat{p}_\theta^2 + \frac{\hat{p}_\phi^2}{\sinh^2\theta}\,.
    \end{align}
\end{subequations}
The first expression, $\hat{p}^2_{\Sigma_1}$,  corresponds to the squared magnitude of the momentum on the two-sphere $\mathbb{S}^2$, first derived in Refs.~\cite{Denicol:2014xca,Denicol:2014tha}. The second, $\pp^2_{\Sigma_0}$, represents the magnitude of the radial momentum in the transverse plane orthogonal to the longitudinal beam direction when written in polar coordinates—after identifying $\theta=r$~\cite{Grozdanov:2025cfx}—and was first obtained by Baym~\cite{Baym:1984np}.  
Finally, the third invariant, $\hat{p}^2_{\Sigma_{-1}}$, which is a novel result first derived in this work, measures the conserved norm of the momentum with respect to the hyperbolic metric $d\Sigma_{-1}^2=d\theta^2+\sinh^2\theta\,d\phi^2$, and therefore defines the invariant squared distance along geodesics of constant negative curvature.  
In this sense, it constitutes the hyperbolic analogue of the Euclidean momentum magnitude. The physical interpretation of the Casimirs~\eqref{eq:mom-cas} will be revisited in Sec.~\ref{subsec:cas_mink}, where we map them back to Minkowski space $\mathbb{R}^{3,1}$.

\subsubsection{Casimir for $E_1$} 
\label{subsub:casE1}
The full isometry group of the spacetime $dS_3\times\mathbb{R}$ includes an additional factor
$E_1 = T(1)\rtimes O(1)$,
whose Lie algebra $\mathfrak{e}_1\cong\mathbb{R}$ has only one Casimir.
For the connected translation subgroup $T(1)$, the Casimir is the conserved translation momentum map $\hat{\mathcal{J}}_{\varsigma} = \hat{p}_\varsigma$.
Under the full $E_1$ action, which includes reflections, the globally invariant quantity is its even combination,
\begin{equation}
\hat{\mathcal{C}}_{\varsigma} = \big(\hat{\mathcal{J}}_{\varsigma}\big)^2 = \hat{p}_\varsigma^2\,.
\label{eq:cas-momlong}
\end{equation}
As expected, this Casimir is the square of the momentum component along the $\varsigma$ direction. 

In summary, when considering the isometry group of a given foliation of \ds, the complete set of global invariants are given by the Casimirs of the foliated $dS_3$, $\hat{C}_\kappa$~\eqref{eq:mom-cas}, and the one associated to $E_1$, 
and the quadratic invariant $\hat{\mathcal{C}}_{\varsigma}$~\eqref{eq:cas-momlong}.

\subsection{Symmetry reduction of the distribution function in \ds}
\label{subsec:dist-theorem}

The Casimirs $\hat{\mathcal{C}}_\kappa$~\eqref{eq:Cas-fol} 
and $\hat{\mathcal{C}}_\varsigma$~\eqref{eq:cas-momlong} constitute the complete set of global invariants associated with the isometry group 
of a given foliation of $dS_3 \times \mathbb{R}$.  
We now apply Proposition~\ref{prop:cohom0} to this geometry, thereby determining the most general functional form of the distribution 
function compatible with the full symmetry of the foliation.

\begin{corollary}[\textit{Reduced functional form of the distribution function on foliated $dS_3\times\mathbb{R}$}]
\label{cor:reduced-form}
Let $f$ be a distribution function on the mass shell 
$\Gamma_m \subset T^*(dS_3 \times \mathbb{R})$, 
invariant under the lifted action of the  isommetry group $H_\kappa \in \{SO(3),\, ISO(2),\, SO(2,1)\}$ and under translation in $\varsigma$,
\begin{equation}
  \{f,\, \hat{\mathcal{J}}^{(\kappa)}_A \} = 0 
  \quad (A=1,2,3)\,,
  \qquad
  \{f,\, \hat{p}_\varsigma \} = 0\,.
\label{eq:invariance-conditions}
\end{equation}
Then, on each connected component of the relevant Liouville-invariant set, 
$f$ can be written as
\begin{equation}
  f = F\!\left(\rho,\; \hat{\mathcal{C}}_\kappa,\; 
  \hat{\mathcal{C}}_\varsigma \right)
  = F\!\left(\rho,\; \hat{p}^{\,2}_{\Sigma_\kappa},\; 
  \hat{p}^{\,2}_\varsigma \right),
\label{eq:reduced-form-dS}
\end{equation}
where $\hat{\mathcal{C}}_\kappa = \hat{p}^{\,2}_{\Sigma_\kappa}$, 
Eq.~\eqref{eq:Cas-fol}, is the quadratic Casimir of the slice 
algebra~$\mathfrak{h}_\kappa \in \{\mathfrak{so}(3),\, \mathfrak{iso}(2),\, \mathfrak{so}(2,1)\}$, and 
$\hat{\mathcal{C}}_\varsigma = \hat{p}^{\,2}_\varsigma$, 
Eq.~\eqref{eq:cas-momlong}, is the Abelian Casimir of~$E_1$.
\end{corollary}
The proof of this corollary is presented in App.~\ref{app:proof-corollary}. 

\subsection{Reduced Boltzmann equation in a \ds foliation}
\label{subsec:redux}

Based on the analysis carried out in the previous section, the isometry group of a foliated \ds demands that 
\begin{equation}
    \label{eq:distrfunc-ds}f^{(\kappa)}(x^\mu,p_i)=f^{(\kappa)}(\rho,\hat{p}^2_{\Sigma_\kappa},\hat{p}_\varsigma^2)\,,
\end{equation}
where $p^2_{\Sigma_\kappa}$ is given by Eq.~\eqref{eq:Cas-fol}. Because the distribution function depends exclusively on the symmetry invariants of the foliated \ds geometry, its derivatives with respect to all coordinates—except $\rho$—and along the momentum directions vanish. The curvature effects of the foliated $dS_3$ are fully encoded in the invariant momentum combinations $\hat{p}^2_{\Sigma_\kappa}$, yielding a single effective Boltzmann equation along $\rho$. Thus, after considering the symmetry arguments, the Boltzmann equation~\eqref{eq:Boltzmann} within the RTA~\eqref{eq:rta} for the foliated \ds reduces to 
\begin{equation}
    \label{eq:redBoltz}
    \frac{\partial}{\partial\rho} f^{(\kappa)}(\rho,\hat{p}^2_{\Sigma_\kappa},\hat{p}_\varsigma^2) = -\frac{\hat{T}_\kappa(\rho)}{c}\left[f^{(\kappa)}(\rho,\hat{p}^2_{\Sigma_\kappa},\hat{p}_\varsigma^2) - f_{eq.}^{(\kappa)}\left(\hat{p}^\rho/\hat{T}_\kappa(\rho)\right)\,\right]\,.
\end{equation}
The resulting reduced equation parallels the kinetic formulations of the Bjorken~\cite{Baym:1984np} and Gubser~\cite{Denicol:2014tha,Denicol:2014xca} boost invariant flows, yet it is now written in a compact and geometrically consistent form within a given foliation leaf $\Sigma_\kappa$ of $dS_3$. Now, in Eq.~\eqref{eq:redBoltz} $\hat{p}^\rho$ is the energy of the particle determined from the on-shell condition 
\begin{equation}
    \label{eq:enerpartds}
    \hat{p}^{\rho} = \sqrt{\frac{\hat{p}^2_{\Sigma_\kappa}}{S^2_\kappa(\rho)}+\hat{p}_\varsigma^2}
\end{equation}
where the functions $S_\kappa(\theta)$ for $\kappa=0,\pm 1$ are listed in Table~\ref{table:ds3defs}. The exact solution of the Boltzmann Eq.~\eqref{eq:redBoltz} reads as
\begin{equation}
f^{(\kappa)}(\rho,\hat{p}^2_{\Sigma_\kappa},\hat{p}_\varsigma)=
D_\kappa(\rho,\rho_0) f^{(\kappa)}_{0}(\rho_0,\hat{p}^2_{\Sigma_\kappa},\hat{p}_\varsigma)
+\frac{1}{c}\int_{\rho_0}^\rho d\rho^{\prime}\,D_\kappa(\rho,\rho^{\prime })\,
\hat{T}_\kappa(\rho^{\prime })\, 
f^{(\kappa)}_{\mathrm{eq}}(\rho^{\prime},\hat{p}^2_{\Sigma_\kappa},\hat{p}_\varsigma) \, .
\label{eq:boltzman_dssol}
\end{equation}
where in the previous expression the damping function $D(\rho_2,\rho_1)$ is given by 
\begin{equation}
D_\kappa(\rho_2,\rho_1)=\exp\!\left(-\int_{\rho_1}^{\rho_2} d\rho''\,
\frac{\hat{T}_\kappa(\rho'')}{c} \right) .  
\label{eq:damp}
\end{equation}
In Eqs.~\eqref{eq:boltzman_dssol} and~\eqref{eq:damp}, $\rho_0$ is the
initial ``time" in \ds space at which 
$f=f^{(\kappa)}_0(\rho_0,\hat{p}^2_{\Sigma_\kappa},\hat{p}_\varsigma)$. For simplicity, we shall assume that $f^{(\kappa)}_0\equiv f_{eq.}$ where $f_{eq.}$ denotes a thermal equilibrium distribution function. This assumption can, however, be straightforwardly relaxed to describe systems far from equilibrium, as done in the conformal Bjorken~\cite{Florkowski:2013lya,Florkowski:2013lza} and Gubser~\cite{Martinez:2017ibh} flows. A detailed analysis of far-from-equilibrium configurations will be presented elsewhere~\cite{MartPlumb}.

The exact solution~\eqref{eq:boltzman_dssol} constitutes the central outcome of our analysis. While it reduces to the known Bjorken~\cite{Baym:1984np} and Gubser~\cite{Denicol:2014tha,Denicol:2014xca} flows for $\kappa=0$ and $\kappa=1$ respectively, it also unveils a new boost-invariant solution for the negatively curved foliation $dS_3$ ($\kappa=-1$). 
\section{From exact kinetics to macroscopic dynamics in \ds slicings}
\label{sec:exactmacro}
Having obtained the exact solution of the Boltzmann equation on foliated \ds , we now turn to its physical implications. In this section, we evaluate the corresponding energy–momentum tensor and extract the macroscopic observables encoded in the distribution function. These results allow us to identify the hydrodynamic regime that emerges from the exact kinetics, as well as the opposite ballistic limit in which the flow becomes free streaming. Together, these analyses provide a complete characterization of the macroscopic dynamics implied by the microscopic solution.
\subsection{Energy–momentum tensor from the exact solution on foliated \ds geometry}
\label{subsec:tmn-ds}
The components of the energy-momentum tensor for the exact solution $f^{(\kappa)}(\rho,\pp_{\Sigma_\kappa}^2,\pp_\varsigma^2)$~\eqref{eq:boltzman_dssol} can be calculated from their definitions~\eqref{eq:tmn-components} as follows
\begin{subequations}
\label{eq:tmn-exact}
\begin{align}
\label{eq:energyds}
\hat{\varepsilon}_\kappa&=\langle\,(-u\cdot p)^2\,\rangle_\kappa\,,\nonumber\\
&=D_\kappa(\rho,\rho_0)\,\hat{I}_{200}(\hat{T}_{0,\kappa};\rho,\rho_0)\,+\,\frac{1}{c}\int_{\rho_0}^\rho\,\,d\rho'\,\hat{T}_\kappa(\rho')\,D_\kappa(\rho,\rho')\,\hat{I}_{200}(\hat{T}_{\kappa}(\rho');\rho,\rho')
\\
\label{eq:shear-ds-comp}
\hat{\pi}^{(\kappa),\varsigma}{}_{\varsigma}\equiv \hat \pi^{(\kappa)}&=\langle\,\pp^{\langle\varsigma}\pp_{\varsigma\rangle}\,\rangle_\kappa\,,\nonumber\\
&=D_\kappa(\rho,\rho_0)\,\left[\hat{I}_{020}(\hat{T}_{0,\kappa};\rho,\rho_0)-\frac{1}{3}\hat{I}_{200}(\hat{T}_{0,\kappa};\rho,\rho_0)\right]\,\nonumber\\
&+\,\frac{1}{c}\int_{\rho_0}^\rho\,\,d\rho'\,\hat{T}_\kappa(\rho')\,D_\kappa(\rho,\rho')\,\left[\hat{I}_{020}(\hat{T}_{\kappa}(\rho');\rho,\rho')-\frac{1}{3}\hat{I}_{200}(\hat{T}_{\kappa}(\rho');\rho,\rho')\right]\,
\\
 \label{eq:shear-ds-phi}
\hat{\pi}^{(\kappa),\phi}{}_{\phi}&=\langle\,\pp^{\langle\phi}\pp_{\phi\rangle}\,\rangle_\kappa=-\frac{\hat{\pi}^{(\kappa)}}{2}\,,\\
\label{eq:shear-ds-th}
\hat{\pi}^{(\kappa),\theta}{}_{\theta}&=\langle\,\pp^{\langle\theta}\pp_{\theta\rangle}\,\rangle_\kappa=-\frac{\hat{\pi}^{(\kappa)}}{2}\,.
\end{align}
\end{subequations}
where in the previous expressions we denote the initial temperature $\hat{T}_{0,\kappa}\equiv\hat{T}_\kappa(\rho_0)$ while introducing the anisotropic integrals $\hat{I}_{nql}=\langle\,\left(\pp^\rho(\rho)\right)^n\,\left(p_\varsigma\right)^l\,\left(\pp_{\Sigma_\kappa}^2/S_\kappa^2(\rho)\right)^q\,\rangle$ whose explicit expressions are presented in App.~\ref{app:integrals}. The system is conformal and thus, the equation of state is $\mathcal{\hat{P}}=\hat{\varepsilon}/3$. In addition, from Eqs.~\eqref{eq:shear-ds-comp}-\eqref{eq:shear-ds-th}, it follows that there is only one single independent non-vanishing component~\footnote{Since the fluid velocity is considered in the local rest frame $\hat{u}_\mu =(-1,0,0,0)$ then $\hat{u}_\mu\hat{\pi}^{\mu\nu}=0$ such that $\hat{\pi}^{\rho\nu} = 0$ $\forall\,\nu$. The isometry group invariance over the constant curvature slice $\Sigma_\kappa$ implies $\hat{\pi}^{ij}=0$ for $i\neq j$.}.
\subsubsection{Landau matching conditions}
\label{subsub:landau}
The effective temperature of the conformal gas follows from the Landau matching condition, which demands equality between the energy density of the system and that of a fictitious equilibrium state. In the Landau frame, this yields $\hat{\varepsilon}_{eq.}=\hat{\varepsilon}$. The equilibrium energy density given by 
\begin{equation}
    \label{eq:energyeq-ds}
    \begin{split}
    \hat{\varepsilon}^{eq.}_{\kappa} (\rho,a)&=
\int \frac{d^3\hat{p}}{(2\pi )^{3}\,S^2_\kappa(\rho)\,Q_\kappa(\theta)\,}\,\hat{p}^\rho\,\,f_{eq.}^{(\kappa)}\left(\hat{p}^{\rho }/\hat{T}(\rho )\right) 
\\
& =\frac{3}{\pi ^{2}}\,\hat{T}_\kappa^{4}(\rho )\,,
    \end{split}
\end{equation}
where the equilibrium distribution function is $f_{eq.}(x)=e^{-x}$. Now,  using Eqs.~\eqref{eq:energyds} and~\eqref{eq:energyeq-ds}, the Landau matching condition implies the following integral equation for the temperature
\begin{equation}
    \label{eq:Landaumatch}
    T_\kappa^4(\rho) = D_\kappa(\rho,\rho_0)\,\hat{R}_{200}\left(\frac{S_\kappa(\rho_0)}{S_\kappa(\rho)}\right)\,\hat{T}^4_{\kappa}(\rho_0)\,+\frac{1}{c}\int_{\rho_0}^\rho\,d\rho'\,\hat{R}_{200}\left(\frac{S_\kappa(\rho')}{S_\kappa(\rho)}\right)\,D_\kappa(\rho,\rho')\,\hat{T}^5_\kappa(\rho')\,.
\end{equation}
The function $\hat{R}_{200}$ is given explicitly in App.~\ref{app:integrals}, Eq.~\eqref{eq:shear-int1}. The integral equation~\eqref{eq:Landaumatch} determines the temperature, which can be obtained numerically using an iterative method~\cite{Banerjee:1989by,Florkowski:2013lza,Florkowski:2013lya,Denicol:2014tha,Denicol:2014xca}. Once the temperature from the above equation and the full distribution function~\eqref{eq:boltzman_dssol} are known, the energy–momentum tensor can be fully determined. The numerical results for foliated geometries including the Grozdanov flow solution~\cite{Grozdanov:2025cfx} will be presented in a forthcoming publication~\cite{MartPlumb}.
\subsection{Emergent Hydrodynamics in foliated \ds}
\label{subsec:hydro}
Hydrodynamics arises from microscopic kinetic theory via a coarse-graining process that lifts microscopic dynamics to macroscopic fluid behavior. For the exact solution of the Boltzmann equation in an arbitrary foliation of \ds~\eqref{eq:boltzman_dssol}, one considers the momentum moments of the distribution function and derives the corresponding equations of motion for these moments.~\cite{DeGroot:1980dk}. In this section, we derive the equations of motion of the hydrodynamical variables by closely following the procedure outlined in Refs.~\cite{Denicol:2014tha,Martinez:2017ibh}. 

\subsubsection{Ideal hydrodynamics}
\label{subsub:ideal}

For the energy density $\hat \varepsilon_\kappa =\langle\,\left(-\hat{u}\cdot\pp\right)^2\,\rangle$ one finds the following evolution equation
\begin{equation}
    \label{eq:energyds-eqn}
    \frac{d\hat{\varepsilon}_\kappa}{d\rho}=\frac{d}{d\rho}\left(\langle(-u\cdot\pp)^2\rangle_\kappa\right)\quad\Rightarrow\quad
    \frac{d\ln{\hat \varepsilon_\kappa}}{d\rho}\,+\,R_\kappa(\rho)\,\left(\frac{8}{3}- \frac{\hat \pi^{(\kappa)}}{\hat \varepsilon_\kappa}\right)=0\,,
\end{equation}
with $R_\kappa(\rho)=\left(dS_\kappa(\rho)/d\rho\right)/S_\kappa(\rho)$. 
The ideal hydrodynamical case occurs when $\eta/\mathcal{S}\to 0$ $(c\to 0)$ so the shear stress tensor is negligible. In this limiting case, the general solution to the ideal hydrodynamical equation~\eqref{eq:energyds-eqn} reads as
\begin{equation}
    \label{eq:idealhydro_sol}
    \hat{\varepsilon}_\kappa(\rho) = \frac{\hat{\varepsilon}_\kappa(\rho_0)}{\left[S_\kappa(\rho)\right]^{8/3}}\,=
    \begin{cases}
        \frac{\hat{\varepsilon}_1(\rho_0)}{\left[\cosh\rho\right]^{8/3}}&\kappa=1 \,,\\
         \hat{\varepsilon}_0(\rho_0)\,e^{8/3\rho}& \kappa=0\,,\\
        \frac{\hat{\varepsilon}_{-1}(\rho_0)}{\left[\sinh\rho\right]^{8/3}}&\kappa=-1\,,
    \end{cases}
\end{equation}
where we used explicitly the functions $S_\kappa$ listed in Table~\ref{table:ds3defs}.
These results were found previously by Grozdanov~\cite{Grozdanov:2025cfx}. Implementing the Weyl scaling to map back to Minkowski space, and applying the coordinate transformation listed in Table~\ref{table:minkdefs}, one obtains Eqs.~(2.18)–(2.20) of Ref.~\cite{Grozdanov:2025cfx}.
\subsubsection{Viscous hydrodynamics}
\label{subsub:2ndhydro}
The evolution equation for the shear viscous stress tensor $\hat{\pi}^{\mu\nu}_\kappa=\langle\,\pp^{\langle\mu}\,\pp^{\nu\rangle}\,\rangle_\kappa$ can also be derived from its definition~\eqref{eq:shear}. In this case, and following Ref.~\cite{Denicol:2014tha}, it is convenient to rewrite the original Boltzmann equation on foliated \ds~\eqref{eq:redBoltz} for the deviation from equilibrium $\delta f=f^{(\kappa)}-f_{eq.}^{(\kappa)}$ as follows 
\begin{equation}
    \label{eq:Boltzmanneq-deltaf}
    \frac{\partial\delta f_\kappa}{\partial\rho}  =-\frac{\delta f_\kappa}{\hat \tau_r} -\frac{\partial f_{eq.}^{(\kappa)}}{\partial\rho}\,.
\end{equation}
Then, differentiating $\hat \pi^{\mu\nu}_\kappa$ with respect to $\rho$ and making use of \eqref{eq:Boltzmanneq-deltaf}, one obtains the following evolution equation\footnote{The Landau matching conditions imply that $\langle\,\pp^{\langle\mu}\,\pp^{\nu\rangle}\,\rangle_{eq.}=0$.}
\begin{equation}
    \label{eq:shearvisc-eqn}
    \begin{split}
        \partial_\rho\hat \pi^{\mu\nu}_\kappa &=\hat \Delta^{\mu\nu}_{(\kappa),\alpha\beta}\frac{\partial}{\partial\rho}\left(\frac{1}{(2\pi)^3}\int\frac{d^3p}{S_\kappa^2(\rho)\,Q_\kappa(\theta)}\,\pp^{\langle\alpha}\pp^{\beta\rangle}\right)\,,
        \\
        &=- \frac{\pi^{\mu\nu}_\kappa}{\tau_r} -2\,R_\kappa(\rho) \pi^{\mu\nu}_\kappa-\frac{R_\kappa(\rho)}{\hat T}\int_\pp \frac{\pp^{\langle\mu}\pp^{\nu\rangle}}{\pp^\rho}\,\left(\frac{\pp^2_{\Sigma_\kappa}}{S_\kappa^2}\right)\,\delta f\\
        &-\int_\pp\,\pp^{\langle\mu}\pp^{\nu\rangle}\left(\frac{\partial_\rho \pp^\rho}{\pp^\rho}\right)\,\delta f\,.
    \end{split}
\end{equation}
It is sufficient to consider the $\varsigma\varsigma$ component of the shear stress tensor $\hat{\pi}^{\varsigma\varsigma}_\kappa\equiv\hat{\pi}^{(\kappa)}$ as it was demonstrated in Sect.~\ref{subsec:tmn-ds}, Eqs.~\eqref{eq:shear-ds-comp}-\eqref{eq:shear-ds-th}. After a bit of algebra and performing some momentum integrals, one finally obtains the following integro-differential evolution equation:
\begin{equation}
    \label{eq:shearsingle-eqn}
    \begin{split}
        \partial_\rho \hat{\pi}^{(\kappa)} =&-\frac{\hat{\pi}^{(\kappa)}}{\hat \tau_r}\,+\,\frac{4}{3}\frac{\eta}{\hat \tau_r}\,R_\kappa\,-\,\frac{46}{21}R_\kappa\,\hat{\pi}^{(\kappa)}\\
        &+\,\frac{R_\kappa}{3(2\pi)}\,\int_0^\infty\,d\pp^\rho\int_0^{2\pi}\,d\theta (p^\rho)^3\sin\theta\left(\frac{25}{21}-\cos^2\theta \right)\,\left(3\cos^2\theta-1\right)\,\delta f\,.
    \end{split}
\end{equation}
We emphasize that the result \eqref{eq:shearsingle-eqn} is exact up to second-order contributions and hence necessarily differs from more standard second-order theories such as Israel–Stewart (IS) approaches \cite{Muller:1967zza,Israel:1979wp} in which certain second-order contributions may be discarded when approximating the distribution function (see App.~A of Ref.~\cite{Denicol:2014tha}).  In terms of the complete result Eq.~\eqref{eq:shearsingle-eqn}, we may then identify two distinct regimes: the Navier-Stokes limit and the limit of second-order conformal hydrodynamics.

\subsubsection{Navier-Stokes limit} 
This regime follows from Eq.~\eqref{eq:shearsingle-eqn} by taking the limit $\hat{\tau}_r\to 0$. Consequently, the NS shear viscous tensor reduces to
    \begin{equation}
    \label{eq:NS-limit}
    \hat{\pi}^{(\kappa)}_{NS}= \frac{4}{3}\eta\,\,R_\kappa = 
    \begin{cases}
        \frac{4}{3}\eta\,\tanh\rho & \kappa = +1\,,\\
        -\frac{4}{3}\eta & \kappa = 0 \,,\\
        \frac{4}{3}\eta\,\coth\rho & \kappa = -1 \,.
    \end{cases}
\end{equation}
Where we used the functions $S_\kappa$ listed in Table~\ref{table:ds3defs}. The NS shear viscous expressions~\eqref{eq:NS-limit} coincide with the results derived by Grozdanov~\cite{Grozdanov:2025cfx}~\footnote{The comparison between Eqs.~\eqref{eq:NS-limit} and the NS solution by Grozdanov~\cite{Grozdanov:2025cfx} is immediate once one recognizes that his result is written in terms of $\hat \pi^{\mu\nu}\sim \eta \,\hat \sigma^{\mu\nu}$ with $\eta=H_0\,\hat{\varepsilon}^{3/4}$.}.

\subsubsection{Second-order conformal hydrodynamics}
If one discards the second-line contributions in Eq.~\eqref{eq:shearsingle-eqn}, i.e., the corrections associated with the exact kinetic treatment, it leads directly to the full second-order conformal equation for the $\hat \pi^{\varsigma\varsigma}_\kappa$
\begin{equation}
    \label{eq:twoorder-eq}
    \partial_\rho\hat \pi^{\varsigma\varsigma}_\kappa =-\frac{\hat \pi^{\varsigma\varsigma}_\kappa}{\hat \tau_r}\,+\,\frac{4}{3}\frac{\eta}{\hat \tau_r}\,R_\kappa\,-\,\frac{46}{21}R_\kappa\,\hat \pi^{\varsigma\varsigma}_\kappa\,.
\end{equation}
It is convenient to solve the second-order differential equations for the Weyl invariant effective shear component $\bar \pi_\kappa=(3/4)\,\hat \pi^{(\kappa)}/\hat{\varepsilon}_\kappa$. Therefore, the evolution equations of second order hydrodynamics read as
\begin{subequations}
\label{eq:2ndordervischydro}
    \begin{align}
    \label{eq:energyresc-eq}
    &\frac{d\ln\hat \varepsilon_\kappa}{d\rho}\,+\,\frac{4\,R_\kappa}{3}\left(2-\bar \pi_\kappa\right)=0\,,\\
    \label{eq:2nd-normpi}
   & \hat \tau_r\left( \frac{d\bar \pi_\kappa}{d\rho} +\frac{4}{3}R_\kappa\,\bar \pi_\kappa^2\right)\,+\,\bar \pi_\kappa\,=\,\frac{\eta}{\hat \varepsilon_\kappa}\,R_\kappa+\frac{10}{21}\,\hat \tau_r\,R_\kappa\,\bar \pi_\kappa\,. 
    \end{align}
\end{subequations}
Neglecting the $(10/21)$ term in Eq.~\eqref{eq:2nd-normpi} reduces this equation to the IS form studied in Ref.~\cite{Marrochio:2013wla} for the Gubser flow and by Soloviev~\cite{Soloviev:2025uig} for the Grozdanov flow. In any foliation of \ds the IS evolution equation of the effective shear is
\begin{equation}
\label{eq:IS-barpi}
\text{IS theory:}\quad \quad \hat \tau_r\left( \frac{d\bar \pi_\kappa}{d\rho} +\frac{4}{3}R_\kappa\,\bar \pi_\kappa^2\right)\,+\,\bar \pi_\kappa\,=\,\frac{\eta}{\hat \varepsilon_\kappa}\,R_\kappa\,.
\end{equation}
The equation for the effective shear~\eqref{eq:2nd-normpi} admits an  analytic solution in the so-called cold limit~\cite{Marrochio:2013wla}. In this regime, where the fluid is highly viscous or extremely cold - namely when $\eta/\left(\mathcal{S} \,T\right)\gg 1$ - the terms proportional to $\hat \tau_r$ dominate in Eq.~\eqref{eq:2nd-normpi}. The effective cold limit equation for the effective shear then reads
\begin{equation}
    \label{eq:coldeqn}
    c\,\left( \frac{d\bar \pi_\kappa}{d\rho} +\frac{4}{3}R_\kappa\,\bar \pi_\kappa^2\right)\,=\,\frac{4}{3}\,R_\kappa+\frac{10}{21}c\,\,R_\kappa\,\bar \pi_\kappa\,,
\end{equation}
where we used $\eta/\hat{\varepsilon}_\kappa = (4/3)\,\hat \tau_r/c$ being $c$ a constant. In the cold limit, the evolution of the shear viscous tensor reduces to a 
Riccati-type differential equation. The crucial difference is that 
Eq.~\eqref{eq:coldeqn} corresponds to a nonlinear \emph{hyperbolic} Riccati flow, whereas the IS counterparts in 
Refs.~\cite{Marrochio:2013wla,Soloviev:2025uig} fall into the 
\emph{parabolic} class.\footnote{The terminology ‘parabolic’ and ‘hyperbolic’ for Riccati equations refers not to an ODE classification but to the geometric classification of one-parameter subgroups of $SL(2,\mathbb{R})$ acting projectively on $\mathbb{RP}^1$. Any Riccati equation with constant coefficients, $\pi' = a + b\pi + c\pi^2$, is the projection of a linear flow $\dot\Psi = X\Psi$ with $X\in\mathfrak{sl}(2,\mathbb{R})$, and its solution is a Möbius transformation generated by $\exp(\rho X)\in SL(2,\mathbb{R})$~\cite{Carinena:1998pun}. Elements of $SL(2,\mathbb{R})$ are classified by their trace: elliptic ($|\mathrm{Tr}|<2$), parabolic ($|\mathrm{Tr}|=2$), and hyperbolic ($|\mathrm{Tr}|>2$)~\cite{kisil2012geometry}.} For any foliation of \ds, the resulting cold-limit approximate solution to Eq.~\eqref{eq:coldeqn} and the IS counterpart, $\bar \pi^\mathrm{cold}_\kappa$ and $\bar \pi^\mathrm{IS-cold}_\kappa$, are given by
\begin{subequations}
\label{eq:gencoldsols}
\begin{align}
\bar \pi_{\kappa}^\mathrm{cold}(\rho)&=
\frac{\bar \pi_+ - \bar \pi_0\,\bar \pi_-\,[S_\kappa(\rho)]^{-\frac{4}{3}(\bar \pi_+-\bar \pi_-)}}
     {1 - \bar \pi_0\,[S_\kappa(\rho)]^{-\frac{4}{3}(\bar \pi_+-\bar \pi_-)}}\,,
     \label{eq:cold-sol}
     \\
     \label{eq:IS-cold}
     \bar \pi_\kappa^\mathrm{IS-cold}(\rho) &= \frac{1}{\sqrt{c}}\, \tanh\left(\frac{4}{3\sqrt{c}}\left[\ln S_\kappa(\rho)-\bar \pi_0\,c\right]\right)\,,
\end{align}
\end{subequations}
where $\bar \pi_0$ is an integration constant and the $c$-dependent fixed points of Eq.~\eqref{eq:coldeqn}, $\bar \pi_\pm$, are
\begin{equation}
\bar \pi_{\pm}(c)
= \frac{5}{28}\pm \frac{1}{28}\sqrt{25+\frac{784}{c}}\,.
\label{eq:fixedpoints}
\end{equation}
The derivation of the cold limit solutions~\eqref{eq:gencoldsols} is presented in App.~\ref{app:coldODE}. By inserting any of the cold-limit solutions for the effective shear~\eqref{eq:gencoldsols} into Eq.~\eqref{eq:energyresc-eq}, one obtains the corresponding cold-limit solutions for the energy density, namely~\footnote{Since $c>0$, the cold-limit fixed points~\eqref{eq:fixedpoints} satisfy
$\pi_{+}>0$ and $\pi_{-}<0$, with $\pi_{+}-\pi_{-}=\mathcal{O}(1)$ for
typical values $c\sim3$--$10$.  In the cold-limit energy-density
solution~\eqref{eq:energyds-coldsol}, the exponent $(4/3)(\pi_{-}-2)$ is
strictly negative, so the leading behavior always decays with
$S_\kappa$ rather than blowing up.  Thus any zero in the bracket
$1-\bar\pi_0^{-1}S_\kappa^{\frac{4}{3}(\pi_{+}-\pi_{-})}$ reflects only
the choice of integration constant or the breakdown of the cold-limit
approximation, not a physical divergence.
}
\begin{subequations}
\label{eq:energycoldgensols}
\begin{align}
    \label{eq:energyds-coldsol}
    \hat \varepsilon_\kappa^\mathrm{cold}(\rho)&= \hat{\varepsilon}_{\kappa}(\rho_0)\,S_\kappa(\rho)^{\frac{4}{3}(\bar \pi_- -2)}\left[1-\bar \pi_0^{-1}\,\left(S_\kappa(\rho)\right)^{\frac{4}{3}\left(\bar \pi_+-\bar \pi_-\right)}\right]\,,\\
    \label{eq:energyds-coldISsol}
    \hat \varepsilon_\kappa^\mathrm{IS-cold}(\rho)&= \hat{\varepsilon}_{\kappa}(\rho_0)\,S_\kappa(\rho)^{-8/3}\,\cosh\left(\frac{4}{3\sqrt{c}}\left[\ln S_\kappa(\rho)-\bar \pi_0\,c\right]\right)
    \,.
    \end{align}
\end{subequations}
%
%\iffalse
\begin{figure}
\centering
\begin{tabular}[c]{lr}
\includegraphics[keepaspectratio,height=\PanelHeight]{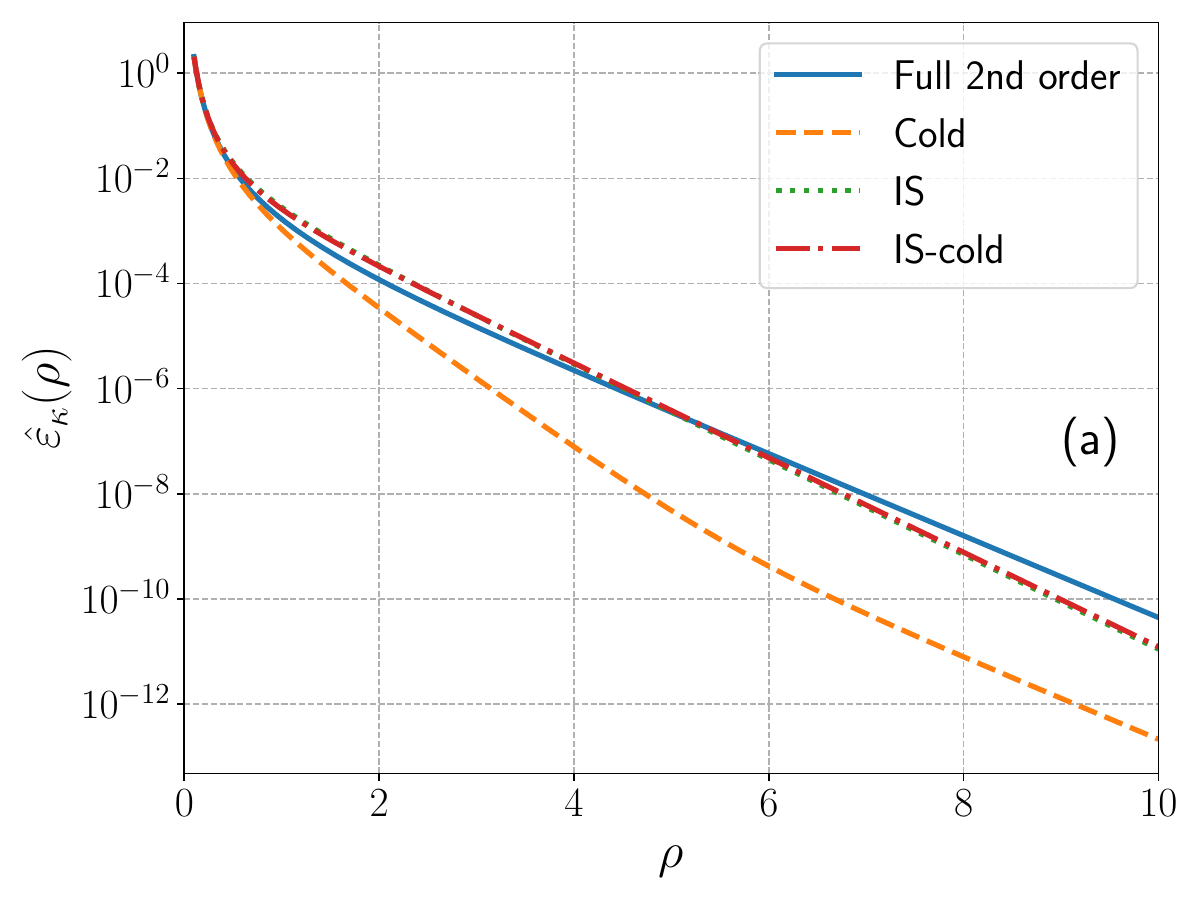}&
\includegraphics[keepaspectratio,height=\PanelHeight]{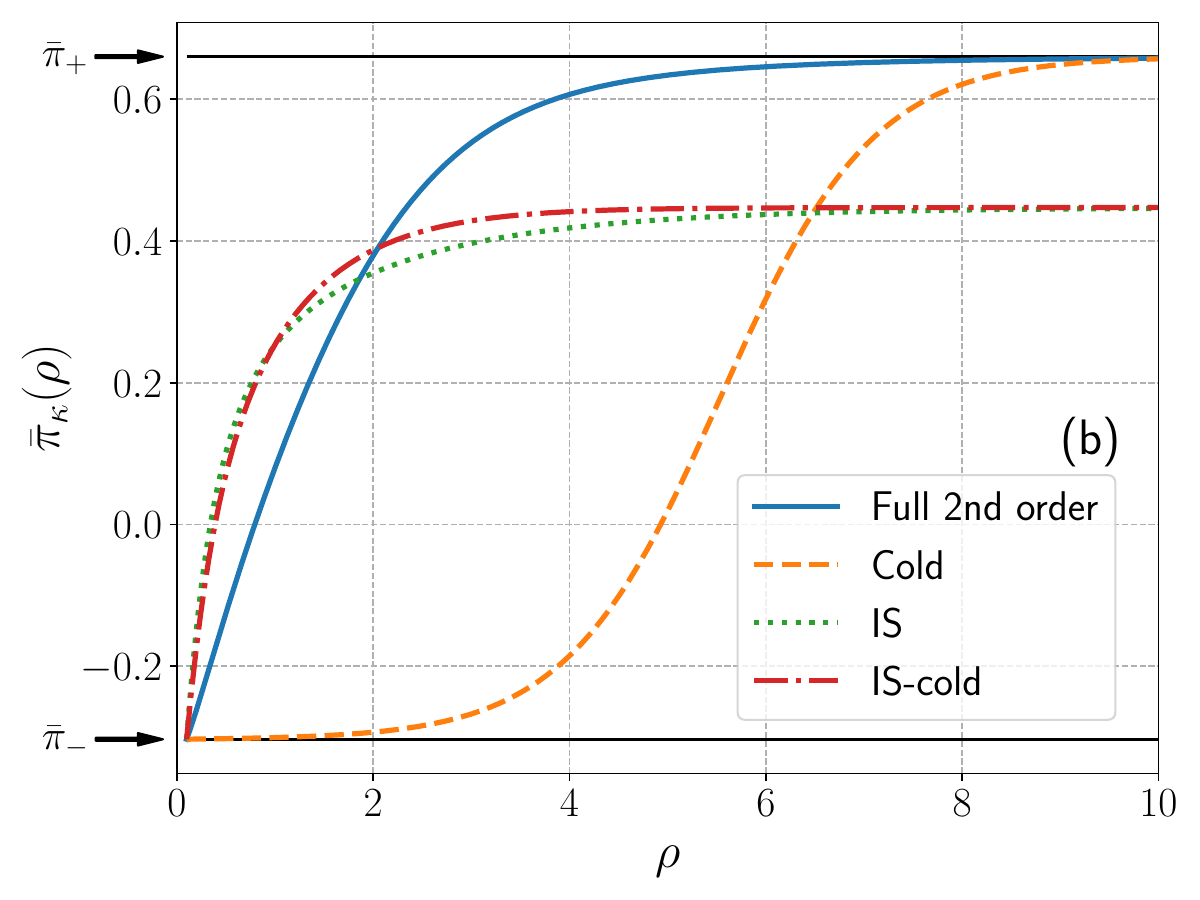}
\end{tabular}
\caption{Comparison of viscous hydrodynamical evolution schemes for the Grozdanov flow. Panel (a): Energy density $\hat \varepsilon_{-1}(\rho)$ and panel (b) effective shear $\bar \pi_{-1}(\rho)$ for four dynamical schemes: the full second-order hydrodynamic equation (solid blue line), Eqs.~\eqref{eq:2ndordervischydro} , its cold-limit approximation (orange long dashed), Eqs.~\eqref{eq:cold-sol}-\eqref{eq:energyds-coldsol}, the Israel–Stewart (IS) equation~(green dotted), Eq.~\eqref{eq:IS-barpi}, and the IS cold-limit solution (red dot-dashed), Eqs.\eqref{eq:IS-cold}-~\eqref{eq:energyds-coldISsol}. Panel (b) displays the relaxation of the effective shear panel (a) highlights the distinct decay rates of the energy density associated with each approximation. The initial conditions for the energy density where $\hat \varepsilon_{-1}(\rho)$ while for the effective shear $\bar \pi_{-1}(\rho_0)=\bar \pi_-(c)$ (where $\bar \pi_-(c)$ is given by Eq.~\eqref{eq:fixedpoints}) with $\rho_0=0.1$, $c=5$ and $\eta/S=0.2$. }
\label{fig:2ndorder}
\end{figure}
%\fi
%
In Fig.~\ref{fig:2ndorder} we show the evolution of the energy density and effective shear in panels (a) and (b), respectively, for second-order hydrodynamical regimes under different approximations; a detailed description is provided in the caption. The cold limit solutions reveal a clear and unified physical picture for 
the energy density and effective shear stress across all \ds foliations. In this regime, the evolution equations simplify to Riccati - type flows whose structure is dictated by the geometry of the foliation through the scale factor $S_\kappa(\rho)$. Both the full second order and the IS truncations produce 
shear solutions of the form $\bar\pi_\kappa(\rho)=\bar\pi_\kappa(S_\kappa(\rho))$, with 
the energy density inheriting the same functional dependence once the shear 
is substituted into the conservation law. As a consequence, the shear and the 
energy density exhibit parallel power law scalings governed by the expanding factor $S_\kappa(\rho)$ and, in the full second order case, by the fixed points $\bar\pi_\pm$ of the cold limit equation~\eqref{eq:coldeqn}. This yields a coherent 
dynamical picture: the shear relaxes toward an asymptotic forward attractor (IS) or 
between two fixed points (full second order) as illustrated in panel (b) of  Figure~\ref{fig:2ndorder}, while the energy density decays monotonically with expansion in a manner entirely determined by the same underlying geometric and viscous parameters as shown in panel (a) of Figure~\ref{fig:2ndorder}.

We conclude this section by highlighting several features of the energy density (or temperature) evolution once viscous corrections are included in any \ds\ foliation. For clarity, we focus on the cold-limit regime of second-order hydrodynamics, although the same conclusions extend to the full linear and nonlinear theories. A key distinction between NS and second-order hydrodynamical theories becomes evident when comparing the shear-stress and energy-density solutions in the Gubser and Grozdanov flows. In NS theory, the absence of a relaxation equation forces the shear to follow algebraically from gradients, leaving no mechanism to dynamically regulate large viscous corrections. For the 
maximally symmetric flows of Gubser and Grozdanov, this leads to unphysical 
negative temperatures in different $|\rho|$ regions. In second order hydrodynamics, by contrast, the cold limit Riccati evolution ensures that the shear remains bounded for appropriate initial conditions, which in turn 
guarantees physically consistent and positive energy density and temperature 
profiles as shown in Fig.~\ref{fig:2ndorder}. In the IS truncation, where the linear term in the Riccati equation 
is absent, the resulting shear flow is parabolic and exhibits a 
$\tanh$--type relaxation, Eq.~\eqref{eq:IS-cold}, toward its forward attractor. 
When the linear term is retained, the full second order theory produces a 
hyperbolic Riccati flow interpolating between two fixed points through a 
M\"obius transformation~\footnote{A M\"obius, or fractional linear, transformation is a map of the form 
\begin{equation*}
    f(x) = \frac{A\,x\,+\,B}{C\,x\,+\,D}
\end{equation*}
with $AD-BC\neq 0$. In the cold limit solution of Eq.~\eqref{eq:cold-sol} this structure appears with $x=\left(S_\kappa(\rho)\right)^{-\frac{4}{3}(\bar \pi_+-\bar \pi_-)}$.
}. In both cases, the bounded nature of the shear prevents the NS pathology, a feature also reflected in the improved behavior of Gubser flow within the causal first order NS theory of Bemfica, Disconzi, and Noronha~\cite{Bemfica:2017wps}. In this formulation, the evolution equations remain regular and avoid the emergence of negative temperatures - an indication of the breakdown of the Landau frame - while still reproducing the qualitative IS-like regulation of dissipative corrections.

\subsection{Free streaming in \ds slices}
\label{sub:free_streaming}
A particularly relevant regime in kinetic theory occurs when the expansion rate exceeds the collision rate, driving the system into the non-hydrodynamic free-streaming propagation. In the RTA model, this limit is realized by assuming an infinitely long relaxation time, effectively taking $\eta/\mathcal{S}\to\infty \,(c\to\infty)$ so only the first term in the RHS of Eq.~\eqref{eq:boltzman_dssol} contributes. In this regime, the energy density \eqref{eq:energyds} and $\varsigma\varsigma$ component of the shear stress \eqref{eq:shear-ds-comp} simplify to the following expressions
\begin{subequations}
    \label{eq:fs-limit}
    \begin{align}
        \label{eq:energy-ds-fs}
        \hat{\varepsilon}_{\kappa}^{f.s.} &=\hat{I}_{200}(\hat{T}_{0,\kappa};\rho,\rho_0)\,,\nonumber\\
        \\
        \label{eq:shear-ds-fs}
        \hat{\pi}^{(\kappa)}_{f.s.}&=\left[\hat{I}_{020}(\hat{T}_{0,\kappa};\rho,\rho_0)-\frac{1}{3}\hat{I}_{200}(\hat{T}_{0,\kappa};\rho,\rho_0)\right]\,.
    \end{align}
\end{subequations}
The numerical studies of the free streaming regime and its validity range for a given foliation of \ds will be discussed in a forthcoming publication~\cite{MartPlumb}.
\section{From foliated \ds to Minkowski $\mathbb{R}^{3,1}$}
\label{sec:dstoMink}

\begin{table}[h!]
\centering
\caption{Coordinate transformations $\rho(\tau,r)$ and $\theta(\tau,r)$, together with the corresponding Milne coordinate ranges $x^\mu=(\tau,r,\phi,\varsigma)$ in $\mathbb{R}^{3,1}$ for different spatial curvatures according to Ref.~\cite{Grozdanov:2025cfx}. In all the cases $0 < r < \infty$, $0 \leq \phi < 2\pi$ and $-\infty<\varsigma<\infty$.}
\label{table:minkdefs}
\begin{tabular}{cccc}
\toprule
$\kappa$ & $\rho(\tau,r)$ & $\theta(\tau,r)$ & $\tau$ range  \\
\midrule
$0$   & $\ln{\tau}$  & $r$ & $0 < \tau < \infty$\\
$+1$  & $\text{arcsinh}\left(-\frac{1-q^2(\tau^2-r^2)}{2q\tau}\right)$   & $\arctan\left(\frac{2qr}{1+q^2(\tau^2-r^2)}\right)$ & $0 < \tau < \infty$ \\
$-1$ & $\text{arccosh}\left(\frac{1+q^2(\tau^2-r^2)}{2q\tau}\right)$ & 
$\text{arctanh}\left(\frac{2qr}{1-q^2(\tau^2-r^2)}\right)$ & $\frac{1}{q}<\tau-r$\\
\bottomrule
\end{tabular}
\end{table}
Under a Weyl rescaling, a tensor $Q^{\mu_1\cdots\mu_m}_{\nu_1\cdots\nu_n}$ of rank $(m,n)$ transforms as~\cite{Baier:2007ix} 
\begin{equation}
\label{eq:weyltrans}
Q^{\mu_{1}\cdots\mu_{m}}_{\nu_{1}\cdots\nu_{n}}
\,\longrightarrow\,
\Omega^{\Delta + m - n}(x)
Q^{\mu_{1}\cdots\mu_{m}}_{\nu_{1}\cdots\nu_{n}},
\end{equation}
where $\Delta$ is the canonical dimension and $\Omega(x)$ is the Weyl factor. For all three constant–curvature slicings considered in Sect.~\ref{subsec:Gubser-Yarom}, the conformal map between Minkowski space $\mathbb{R}^{3,1}$ and a given \ds foliation is implemented by the Weyl rescaling 
\begin{equation} 
d\hat s_{\text{\ds}}^2 = \frac{1}{\tau^2} ds_{\mathbb{R}^{3,1}}^2,
\end{equation}
from which we identify the Weyl factor as $\Omega(\tau,r)=\tau^{-1}$. In addition to this scaling, the full conformal map to a particular \ds slice requires a coordinate transformation of the form \( \hat x^\mu = \hat x^\mu(\tau,r;\kappa)
\). The combined Weyl and coordinate transformations, together with the corresponding coordinate ranges for each slicing, are summarized in Table~\ref{table:minkdefs}~\cite{Grozdanov:2025cfx}.

%\iffalse
\begin{figure}%[h]
  \centering
  \begin{tabular}[c]{lr}
    % (left bottom right top)
    \includegraphics[keepaspectratio, height=\PanelHeight]{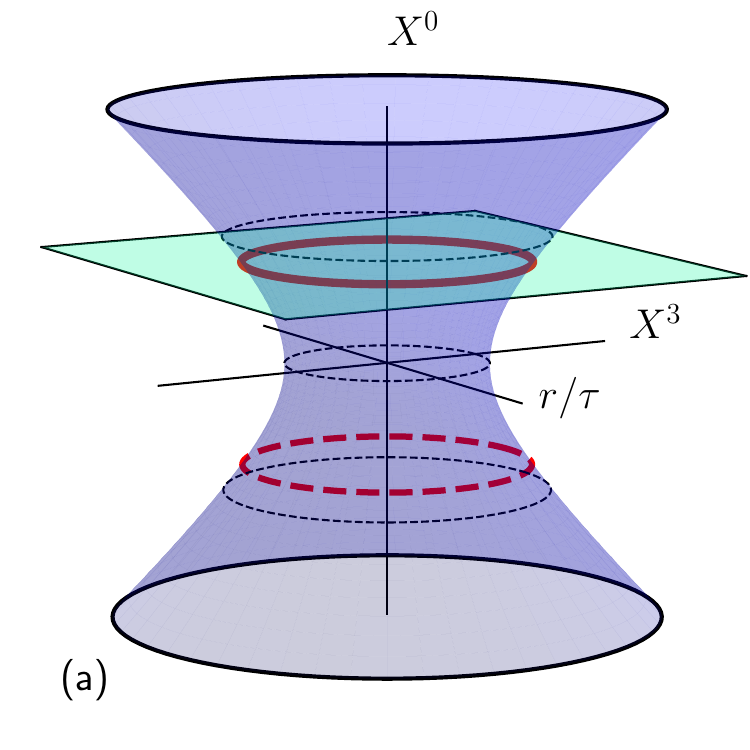}&
    \includegraphics[keepaspectratio, height=\PanelHeight]{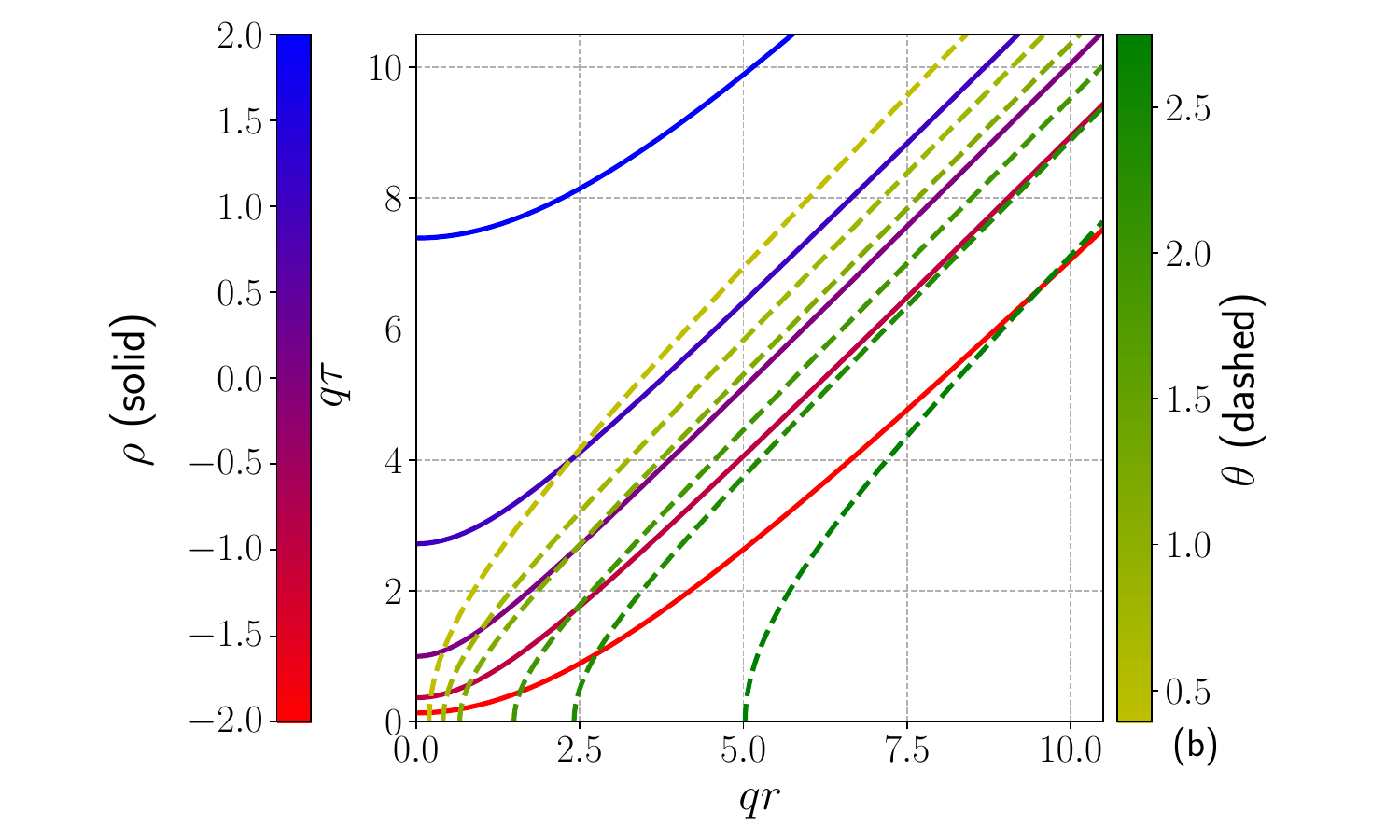}\\
    \includegraphics[keepaspectratio, height=\PanelHeight]{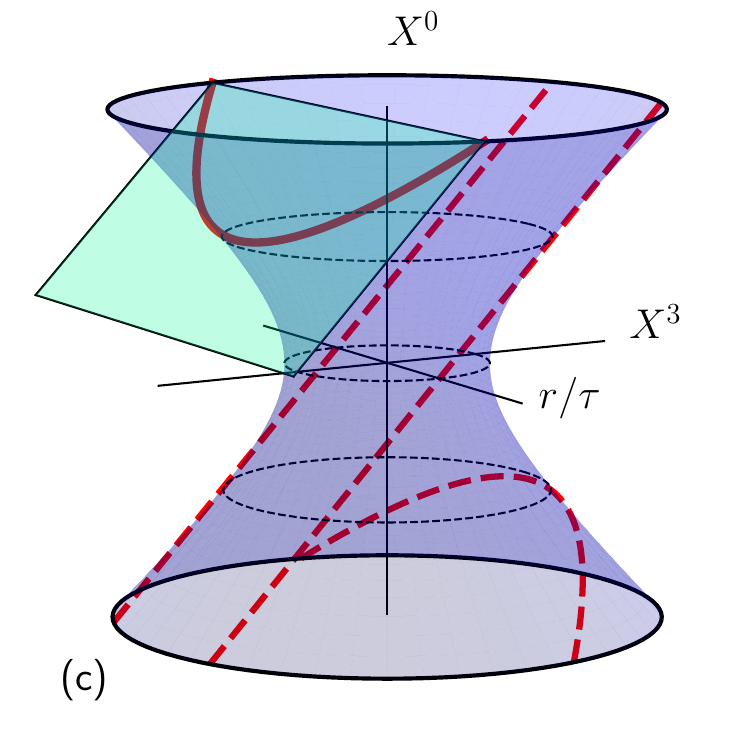}&
    \includegraphics[keepaspectratio, height=\PanelHeight]{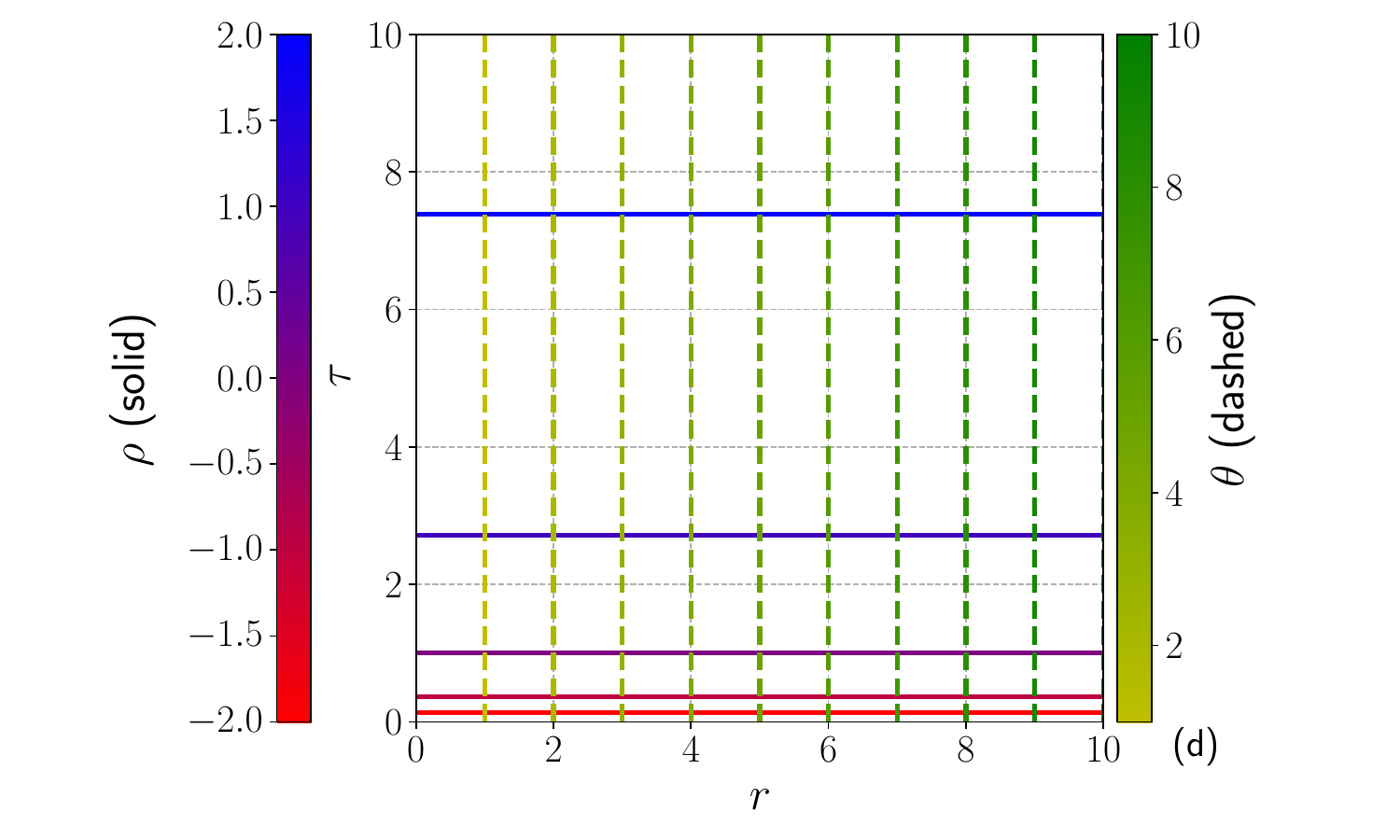}\\
    \includegraphics[keepaspectratio, height=\PanelHeight]{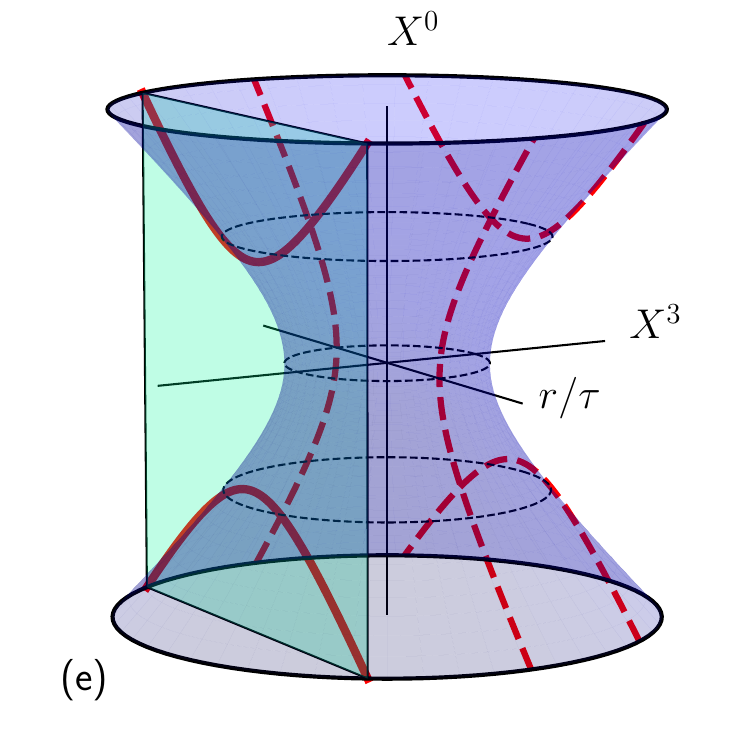}&
    \includegraphics[keepaspectratio, height=\PanelHeight]{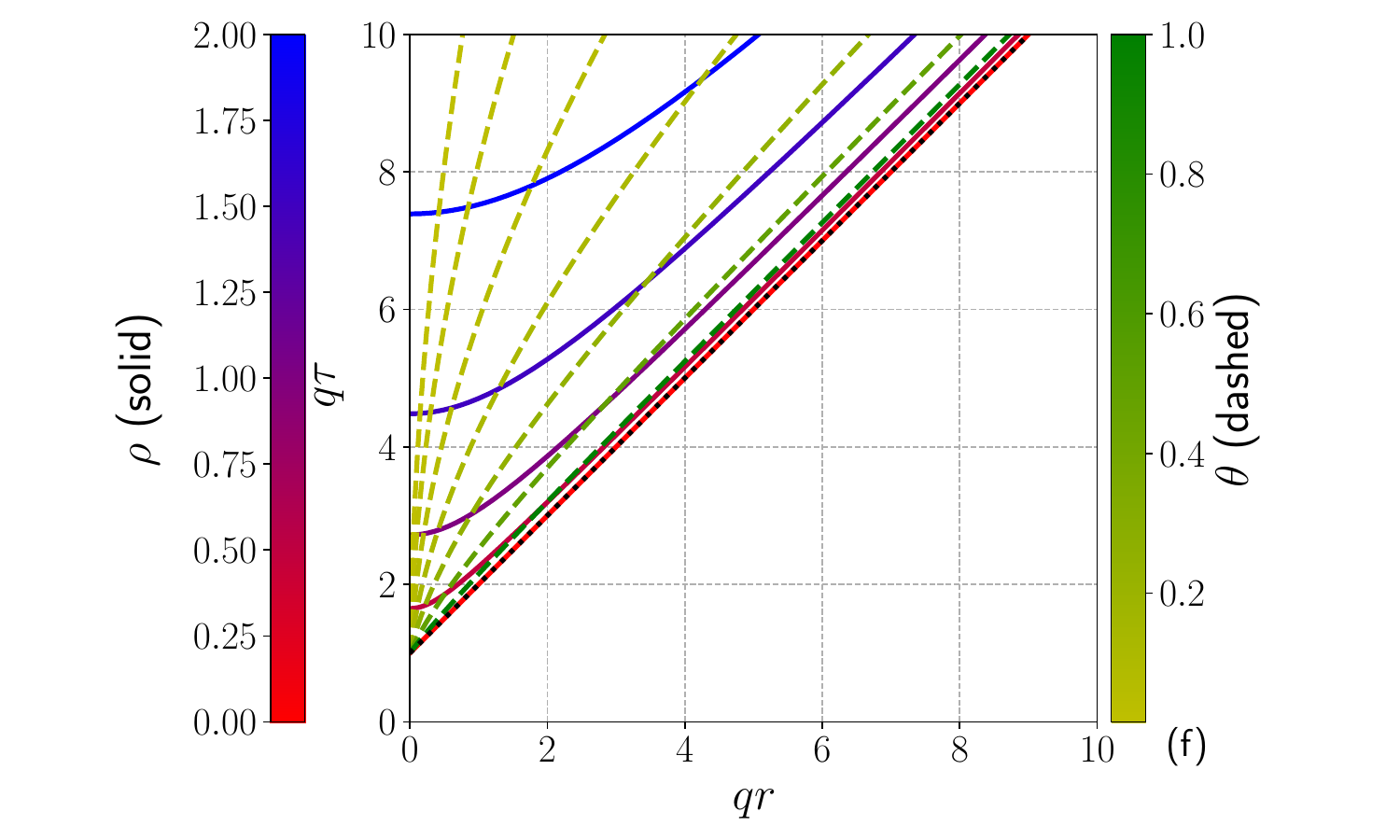}
  \end{tabular}    
  \caption{Different choices of foliation of \ds, parameterized by $(\rho,\theta)$, and the corresponding relations to Milne coordinates $(\tau,r)$.
  \textbf{Left panels}: The Gubser solution (top row; $\kappa = +1$) corresponds to foliations at $X^0 = \text{constant}$; the Bjorken solution (middle row; $\kappa = 0$) is defined by foliating with $X^3 - X^0 = \text{constant}$; and the Grozdanov solution (bottom row; $\kappa = -1$) reflects foliations with respect to $X^3 = \text{constant}$.  Intersections between the chosen level and the hyperboloid are displayed as solid red contours, while other levels are displayed as dashed red contours.  Black dashed contours at constant $X^0$ are included to help visualize the hyperboloid.
  \textbf{Right panels:} relations between $(\rho,\theta)$ and $(\tau,r)$ for each foliation.  The ranges shown for $\rho$ and $\theta$ in each panel are chosen solely for aesthetic purposes; the full ranges are given in \ref{table:ds3defs}.
  }
 \label{fig:foliations}
\end{figure}
%\fi

For clarity, we further depict the foliations and their corresponding coordinate relations in Fig.~\ref{fig:foliations}.
In the lefthand panels of Fig.~\ref{fig:foliations}, we show foliations of the de Sitter hyperboloid (blue surface) at constant $X^0$ (Gubser - top), constant $X^3-X^0$ (Bjorken - middle), and constant $X^3$ (Grozdanov - bottom), each with a representative leaf from that foliation (green planes) and its intersection with the hyperboloid (solid red curves).  The dashed red curves similarly display intersections of the hyperboloid with other leaves (not shown) from the same foliation.  One observes that leaves in the Gubser ($\kappa = +1$) foliation yield compact geometries which are parametrized by $0 \leq \theta \leq 2\pi$, whereas the Bjorken ($\kappa = 0$) and Grozdanov ($\kappa = -1$) foliations correspond to open geometries which are either parabolic or hyperbolic, respectively, such that $0 \leq \theta \leq \infty$, consistent with the ranges given in Table \ref{table:ds3defs}.  Each geometry thus reflects the constant curvature $\kappa$ of the corresponding foliation.
In the righthand panels of Fig.~\ref{fig:foliations}, we visualize the relations between the foliated $dS_3$ coordinates $(\rho,\theta)$ and the usual Milne coordinates $(\tau,r)$, with the explicit coordinate transformations provided in Table~\ref{table:minkdefs}.  Whereas the Bjorken coordinates (middle) relate trivially to the corresponding de Sitter coordinates, the Gubser (top) and Grozdanov (bottom) coordinates map non-trivially onto Milne coordinates.  In all foliations, increasing $\rho$ at $r=0$ coincides with increasing $\tau$, while taking $\theta \to 0$ is equivalent to taking $r \to 0$.  Likewise, in the Gubser and Bjorken foliations, one finds that increasing $\theta$ at $\tau = 0$ is equivalent to increasing $r$, whereas the Grozdanov foliation tends toward the limiting line $q\tau = qr + 1$ as $\theta \to +\infty$ or $\rho \to 0$.

\subsection{Interpreting Slice-depending Casimirs in Minkowski Space}
\label{subsec:cas_mink}
In Sec.~\ref{subsub:casds3}, we determined the Casimirs of a given \ds, Eqs.~\eqref{eq:Cas-fol} and~\eqref{eq:mom-cas}, and provided their geometric interpretation. To clarify their physical content, in this section we rewrite these invariants in terms of momenta measured in ordinary Minkowski space. This makes explicit how the conserved quantities on curved slices correspond to boosted and Weyl–rescaled combinations of Minkowski momentum. In this way, the Casimirs provide a unified interpretation of all boost–invariant flows through the geometry of their underlying slicings. We follow closely the approach of Ref.~\cite{Heinz:2015cda}.
\subsubsection{Momentum transformation and radial--boost structure}
\label{subsub:momtransf}
The combination of the Weyl factor and the coordinate transformation between $\mathbb{R}^{3,1}$ and \ds slicings leads to the general momentum transformation
\begin{equation}
    \label{eq:mom-transf}
    \hat p^\mu = \Omega(x)^{-2}\,\left(\frac{\partial\hat x^\mu}{\partial x^\alpha}\right)\,p^\alpha
\end{equation} 
The explicit coordinate mapping is
\begin{align}
\label{eq:boost-transf}
\hat p^\rho &= \tau^2\left(
\frac{\partial\rho}{\partial\tau}\,p^\tau
+ \frac{\partial\rho}{\partial r}\,p^r
\right),\\
\hat p^\theta &= \tau^2\left(
\frac{\partial\theta}{\partial\tau}\,p^\tau
+ \frac{\partial\theta}{\partial r}\,p^r
\right),\\
\hat p^\phi &= \tau^2\,p^\phi,\\
\hat p^\varsigma &= \tau^2\,p^\varsigma.
\end{align}
For the Gubser and Grozdanov flows, the fluid velocity in \ds is static, whereas in Minkowski space $u^\mu$ is purely radial and azimuthally symmetric, so $u^\mu(\tau,r) = \big(u^\tau(\tau,r),\,u^r(\tau,r),\,0,\,0\big)$ which we parametrize as
\begin{equation}
u^\tau = \gamma_\kappa(\tau,r),\qquad
u^r = \gamma_\kappa(\tau,r)\,v_\kappa(\tau,r),
\end{equation}
with
\(
\gamma_\kappa = 1/\sqrt{1-v_\kappa^2}
\). Using the Weyl transformation of vectors~\footnote{Note that the Weyl weights of the fluid velocity $\hat{u}^\mu$ and $\pp^\mu$ differ: the four–velocity is rescaled with $\Omega^{-1}$ so as to preserve the normalization $\hat{u}_\mu \hat{u}^\mu=-1$, whereas the momentum acquires an extra factor $\Omega^{-2}$ because the canonical covector $p_\mu$ is taken as Weyl–invariant and indices are raised with the rescaled metric $\hat{g}^{\mu\nu}=\Omega^{-2}\,g^{\mu\nu}$.}, $\hat u^\mu = \Omega^{-1}\,\left(\partial\hat x^\mu/\partial x^\alpha\,\right)\,u^\alpha$ leads to the following equations
\begin{align}
\tau\,\partial_\tau\rho \;\gamma_\kappa
+ \tau\,\partial_r\rho \;\gamma_\kappa v_\kappa &= 1,\\[2pt]
\tau\,\partial_\tau\theta \;\gamma_\kappa
+ \tau\,\partial_r\theta \;\gamma_\kappa v_\kappa &= 0.
\end{align}
Solving these equations yields the Jacobian derivatives in terms of the radial boost,
\begin{equation}
\label{eq:jacobian-boost}
\begin{aligned}
\tau\,\partial_\tau\rho &= \gamma_\kappa, 
&\qquad \tau\,\partial_r\rho &= -\,\gamma_\kappa v_\kappa,\\[4pt]
\tau\,\partial_\tau\theta &= -\,\frac{\gamma_\kappa v_\kappa}{S_\kappa(\rho)},
&\qquad \tau\,\partial_r\theta &= \frac{\gamma_\kappa}{S_\kappa(\rho)}.
\end{aligned}
\end{equation}
Substituting~\eqref{eq:jacobian-boost} back into the momentum transformation~\eqref{eq:boost-transf} yields to the following general transformation for any constant–curvature slicing ($\kappa=\pm 1$),
\begin{equation}
\label{eq:boosted}
\begin{aligned}
\hat p^\rho &= \tau\,\gamma_\kappa(\tau,r)\,\big(p^\tau - v_\kappa(\tau,r)\,p^r\big),\\[4pt]
\hat p^\theta &= \frac{\gamma_\kappa(\tau,r)}{S_\kappa(\rho(\tau,r))}\,
\big(p^r - v_\kappa(\tau,r)\,p^\tau\big),\\[4pt]
\hat p^\phi &= \tau^2\,p^\phi,\qquad
\hat p^\varsigma = \tau^2\,p^\varsigma.
\end{aligned}
\end{equation}
Therefore, we can understand the particle's momenta in \ds in terms of momenta in the lab frame boosted by a radial flow. 

\subsubsection{Casimirs in terms of boosted momenta}
\label{subsub:casimir-boosted}
Using Eqs.~\eqref{eq:boosted} allows to rewrite the Casimir associated with the constant–curvature slice $\Sigma_\kappa$, Eq.~\eqref{eq:Cas-fol}, in terms of Minkowski momenta as follows
\begin{equation}
\hat{\mathcal{C}}_\kappa 
= \hat p_\theta^2 + \frac{\hat p_\phi^2}{Q_\kappa^2(\theta)}\Rightarrow \mathcal{C}_\kappa 
= \frac{\gamma_\kappa^2}{S_\kappa^2(\rho)}\,
\big(p^r - v_\kappa p^\tau\big)^2
+ \frac{\tau^4 (p^\phi)^2}{Q_\kappa^{2}(\theta)}\,.
\end{equation}
Now, without losing any generality, we shall asume that for  an azimuthally symmetric flow $p^\phi=0$.  For $\kappa=+1$ (spherical slicing, Gubser flow) one has
$S_{+1}(\rho)=\cosh\rho$ and $Q_{+1}(\theta)=\sin\theta$ (see Table~\ref{table:ds3defs}).  the Casimir reduces to
\begin{equation}
\mathcal{C}_{+1}
= \frac{\gamma_{+1}^2}{\cosh^2\rho}\,\big(p^r - v_{+1} p^\tau\big)^2,
\end{equation}
which is equivalent to Eq.~(15) of Ref.~\cite{Heinz:2015cda} after expressing $p^r$ through
the transverse momentum $p_T$. 

For $\kappa=-1$ (hyperbolic slicing, ``Grozdanov flow''), one has that the hyperbolic scale factor $S_{-1}(\rho) = \sinh\rho$, leading to
\begin{equation}
\mathcal{C}_{-1}
= \frac{\gamma_{-1}^2}{\sinh(\rho(\tau,r))}\,
\big(p^r - v_{-1} p^\tau\big)^2
\end{equation}
which represents the invariant momentum norm on the hyperbolic plane $H^2$ -  the geodesic distance in the metric 
$ds^2 = d\theta^2 + \sinh^2\theta\, d\phi^2$, and constitutes the 
characteristic invariant of the Grozdanov flow - written
entirely in terms of Minkowski momenta and the radial velocity field $v_{-1}(\tau,r)$. In our upcoming work~\cite{MartPlumb}, we will numerically assess the physical conditions that delimit the validity and applicability of the novel kinetic solution for the Grozdanov flow.

Notice that for the $\kappa=0$, one can still use the general transformation between momenta in \ds and $\mathbb{R}^{3,1}$, Eqs.~\eqref{eq:boost-transf}, by setting $v_0(\tau,r)=0$. In that case it is straightforward to show that the Casimir $\mathcal{C}_0$ is simply the transverse momentum of the particle, namely
\begin{equation}
\mathcal{C}_0 = p_T^2 =p_r^2+\frac{p_\phi^2}{r^2}\,,
\end{equation}
where we used explicitly the function $Q_0(\theta)=\theta\equiv r$~\cite{Grozdanov:2025cfx}. This result follows purely from the flat $ISO(2)$ symmetry of the transverse plane for the Bjorken flow. 
Finally, the Casimir $\mathcal{C}_\varsigma$ associated to $\mathbb{E}_1$ is identified as the longitudinal momentum which follows directly from Eqs.~\eqref{eq:boosted}.
%=============================
\section{Conclusions}
\label{sec:concl}

In this work we have developed a unified formulation of relativistic kinetic 
theory for an uncharged conformal gas evolving on arbitrary foliations of 
\ds. By exploiting the maximal symmetry of 
$\mathrm{d}S_3$ and its different constant--curvature slicings, we obtained a 
single exact solution of the Boltzmann equation whose restriction to specific 
foliations reproduces the Bjorken~\cite{Baym:1984np} and 
Gubser~\cite{Denicol:2014tha,Denicol:2014xca} \emph{kinetic} solutions, and, in 
addition, yields a new exact kinetic solution describing a gas undergoing 
Grozdanov flow. Thus all three expanding systems arise as different coordinate 
projections of one underlying invariant hyperboloid, exposing a unifying 
geometric origin for a broad class of kinetic solutions that would otherwise 
appear unrelated.

Our formulation makes systematic use of the cotangent bundle description of kinetic theory and its geometric representation in the phase space~\cite{Rioseco:2016jwc,rioseco2019relativistic,Acuna-Cardenas:2021nkj,Sarbach:2013fya,Sarbach:2013hna,Sarbach:2013uba}. This perspective brings to the forefront the roles of symmetry generators, Casimir invariants, 
and momentum maps in constructing exact solutions of the Boltzmann equation. 
It also highlights the intrinsic foliation--independence of the distribution 
function: once lifted to the cotangent bundle, the kinetic solution acquires a 
representation that does not privilege any particular slicing of $\mathrm{d}S_3$. 
Consequently, emergent hydrodynamic behavior---ideal, viscous, or free 
streaming---can be analyzed uniformly across all foliations simply by 
projecting the same exact kinetic solution onto different hypersurfaces. In addition to the exact Boltzmann solution for a system undergoing Grozdanov flow, we have also obtained new approximate solutions for second–order hydrodynamic theories in the so-called cold limit~\cite{Marrochio:2013wla}, now implemented on arbitrary \ds foliations rather than only the Gubser geometry. The cold--limit analysis demonstrates how 
geometric properties of the foliation govern the relaxation of viscous 
stresses and the corresponding evolution of the energy density.

In addition to unifying existing flows, the geometric structure inherent in the cotangent-bundle formulation provides a flexible framework for identifying new exact solutions in systems with suitable symmetries and for probing far–from–equilibrium dynamics. More broadly, this geometric viewpoint opens new avenues for studying non-equilibrium attractors, symmetry-protected transport, and universal features of out-of-equilibrium evolution directly within kinetic theory.

In a forthcoming follow--up work~\cite{MartPlumb}, we will examine the numerical properties 
of the exact solutions of the Boltzmann equation in different foliated geometries for far-from equilibrium situations, with particular emphasis on the structure induced 
by Grozdanov flow and its implications for hydrodynamization, causality and stability.

\acknowledgments{We thank to Sa\v{s}o Grozdanov for useful conversations about his novel hydrodynamical solution~\cite{Grozdanov:2025cfx}. We also thank Olivier Sarbach for helpful correspondence on the cotangent-bundle formulation of relativistic kinetic theory and for a careful reading of the manuscript. His remarks were instrumental in clarifying our original claims and in shaping Proposition 1~\ref{prop:cohom0} and Corollary 1~\ref{cor:reduced-form} in the revised version. M.~M. thanks to C. Taepke-Guerrero for contributing to a creative, harmonious and productive working environment during the completion of this work.  
}
\appendix

\section{Symplectic reduction, cohomogeneity, and proofs}
\label{app:symplectic-reduction}

\subsection{A brief review on Marsden--Weinstein reduction and cohomogeneity}
\label{app:MW-review}

We collect the minimal background needed for the proofs of 
Proposition~\ref{prop:cohom0} and Corollary~\ref{cor:reduced-form}. 
Standard references for symplectic reduction 
are~\cite{MarsdenWeinstein1974,AbrahamMarsden,MarsdenRatiuBook,OrtegaRatiu}. 
For cohomogeneity, 
see~\cite{berndt2017cohomogeneity,diaz2017cohomogeneity}.

\subsubsection{Hamiltonian group actions and momentum maps}
Let $(P,\omega)$ be a symplectic manifold and let a Lie group~$G$ act 
properly on~$P$ by symplectomorphisms. The action is called 
\emph{Hamiltonian} if there exists a map $J:P\to\mathfrak{g}^*$ (the 
\emph{momentum map}) such that for every $\xi\in\mathfrak{g}$,
\begin{equation}
  \iota_{X_\xi}\,\omega = d\langle J,\xi\rangle\,,
  \qquad
  J_\xi := \langle J,\xi\rangle\,,
\label{eq:momentum-map-def}
\end{equation}
where $X_\xi$ is the infinitesimal generator of the action. The momentum 
map is called \emph{equivariant} if 
$J(g\cdot z) = \mathrm{Ad}^*_g\,J(z)$ for all $g\in G$ and $z\in P$.

\subsubsection{Marsden--Weinstein reduction theorem}
Fix $\mu\in\mathfrak{g}^*$ a regular value of~$J$ and let
\begin{equation}
  G_\mu = \bigl\{g\in G \;\big|\; 
  \mathrm{Ad}^*_g\,\mu = \mu\bigr\}
\label{eq:coadjoint-isotropy}
\end{equation}
denote the coadjoint isotropy subgroup. If the $G_\mu$-action on 
$J^{-1}(\mu)$ is free and proper, then the quotient
\begin{equation}
  P_\mu := J^{-1}(\mu)\,/\,G_\mu
\label{eq:reduced-space}
\end{equation}
is a smooth manifold equipped with a unique symplectic 
form~$\omega_\mu$ determined by
\begin{equation}
  \pi^*\omega_\mu = i^*\omega\,,
\label{eq:reduced-symplectic}
\end{equation}
where $i:J^{-1}(\mu)\hookrightarrow P$ is the inclusion and 
$\pi:J^{-1}(\mu)\to P_\mu$ is the projection. Moreover, if 
$H\in C^\infty(P)$ is $G$-invariant, then $H$ descends to a 
Hamiltonian~$H_\mu$ on~$P_\mu$ and the Hamiltonian flow of~$H$ 
restricts to the reduced flow on $(P_\mu,\omega_\mu)$.

The key consequence for kinetic theory is the following: $G$-invariant 
functions on~$P$ factor through the quotient $P/G$. In particular, a 
distribution function satisfying $\{f,J_\xi\}=0$ for all 
$\xi\in\mathfrak{g}$ is constant on $G$-orbits and therefore depends 
only on the invariants that parametrize the orbit space. However, the 
reduced space~$P_\mu$ can itself have positive dimension, in which case 
additional invariants beyond the Casimirs of the coadjoint action---such 
as integrals of the reduced Hamiltonian flow---may appear. An additional 
hypothesis is needed to exclude this possibility.

\subsubsection{Cohomogeneity}
Let $G$ act smoothly on a manifold~$N$. A \emph{principal} (regular) 
orbit is one of maximal dimension; its codimension in~$N$ is constant on 
an open dense subset. The \emph{cohomogeneity} of the action is defined 
as
\begin{equation}
  \mathrm{cohom}(G\curvearrowright N) 
  := \mathrm{codim}_{N}(G\cdot z)\,,
  \qquad z\ \text{regular}\,.
\label{eq:cohomogeneity-def}
\end{equation}
Cohomogeneity zero means the action is transitive on each connected 
component of~$N$, equivalently the orbit space $N/G$ is discrete. In 
that case, any smooth $G$-invariant function on~$N$ is locally constant 
on each connected component, and the only nontrivial functional 
dependence can occur through the labels that distinguish these 
components.

\paragraph{Combined implication.}
When the Marsden--Weinstein reduction is combined with a 
cohomogeneity-zero condition on the relevant level sets, the 
distribution function is forced to depend only on the Casimir invariants 
of~$\mathfrak{g}^*$: the MW reduction ensures that $f$ factors through 
$P/G$, while cohomogeneity zero guarantees that $P/G$ is discrete once 
the Casimirs are fixed, leaving no room for additional continuous 
invariants. This is the content of Proposition~\ref{prop:cohom0}, whose 
proof we now present.

\subsection{Proof of Proposition~\ref{prop:cohom0}}
\label{app:proof-proposition}

The proof proceeds in three steps: factorization through the orbit 
space, restriction to Casimir level sets via cohomogeneity zero, and 
reconstruction as a function of Casimirs.

\medskip\noindent
\textit{Step~1: $G$-invariance implies factorization through the orbit 
space.}
Let $\pi:\mathcal{N}\to\mathcal{N}/G$ denote the quotient map. For 
each~$a$, the infinitesimal generator~$X_a$ of the $G$-action 
on~$\Gamma$ is the Hamiltonian vector field associated 
with~$\mathcal{J}_a$, so the symmetry condition 
$\pounds_{\tilde{\mathcal{K}}_a}\,f = 0$ is equivalent to 
$\{f,\mathcal{J}_a\} = 0$, which in turn gives $df(X_a) = 0$. Since 
the tangent space to the $G$-orbit through any $z\in\mathcal{N}$ is
\begin{equation}
  T_z(G\cdot z) = \mathrm{span}\{X_a(z)\}\,,
\label{eq:orbit-tangent}
\end{equation}
it follows that $f$ is constant on each $G$-orbit. Therefore there 
exists a unique function~$\bar{f}$ on the orbit space such that
\begin{equation}
  f = \bar{f}\circ\pi\,.
\label{eq:f-factors}
\end{equation}

\medskip\noindent
\textit{Step~2: Cohomogeneity zero forces local constancy on Casimir 
level sets.}
Let $\{\mathcal{C}_i\}$ be a complete set of coadjoint invariants 
(Casimirs) on the relevant stratum of~$\mathfrak{g}^*$, and fix 
values~$c_i$. Define the level set
\begin{equation}
  \mathcal{N}_{\mathbf{c}} 
  := \bigl\{z\in\mathcal{N} \;\big|\; 
  \mathcal{C}_i\bigl(J(z)\bigr) = c_i\bigr\}\,.
\label{eq:casimir-level-set}
\end{equation}
By equivariance of the momentum map, 
$J(g\cdot z) = \mathrm{Ad}^*_g\,J(z)$, any coadjoint invariant 
satisfies $\mathcal{C}_i\bigl(J(g\cdot z)\bigr) = 
\mathcal{C}_i\bigl(J(z)\bigr)$, so $\mathcal{N}_{\mathbf{c}}$ is 
$G$-invariant.

By hypothesis, the $G$-action has cohomogeneity zero on each connected 
component of~$\mathcal{N}_{\mathbf{c}}$. This means $G$ acts 
transitively there, so the quotient 
$\mathcal{N}_{\mathbf{c}}/G$ is discrete on each connected component. 
By Step~1, $f$ factors through this quotient; since the quotient is 
discrete, $\bar{f}$ is locally constant on each connected component 
of~$\mathcal{N}_{\mathbf{c}}/G$. In other words, once the Casimir 
values are fixed, $f$ admits no further continuous variation.

\medskip\noindent
\textit{Step~3: Reconstruction as a function of Casimirs.}
Let $z$ and $z'$ lie in the same connected component 
of~$\mathcal{N}$, and suppose 
$\mathcal{C}_i\bigl(J(z)\bigr) = \mathcal{C}_i\bigl(J(z')\bigr)$ for 
all~$i$. Then both $z$ and $z'$ belong to~$\mathcal{N}_{\mathbf{c}}$ 
for the same~$\mathbf{c}$, and by Step~2, $f(z) = f(z')$. Therefore 
the value of~$f$ depends locally only on the collection 
$\{\mathcal{C}_i(J)\}$. This means there exists a smooth function~$F$ 
such that
\begin{equation}
  f = F\!\bigl(\mathcal{C}_1(J),\,\mathcal{C}_2(J),\,\ldots\bigr)
\label{eq:f-casimir-final}
\end{equation}
locally on each connected component, which is 
Eq.~\eqref{eq:invariantsol}.\hfill$\square$

\subsection{Proof of Corollary~\ref{cor:reduced-form}}
\label{app:proof-corollary}

We verify that the hypotheses of Proposition~\ref{prop:cohom0} are 
satisfied for the symmetry group $G:=H_\kappa\times E_1$ acting on the 
mass shell $\Gamma_m\subset T^*(dS_3\times\mathbb{R})$, and then apply 
the proposition to obtain the reduced functional form. The proof has 
four steps.

\medskip\noindent
\textit{Step~1: Hamiltonian action and momentum maps.}
The foliated metric~\eqref{eq:fol-met} admits the isometry group 
$H_\kappa\in\{SO(3),\,ISO(2),\,SO(2,1)\}$ generated by the Killing 
vectors $\hat{\mathcal{K}}^{(\kappa)}_A$ ($A=1,2,3$), 
Eqs.~\eqref{eq:Killvectsds3R}, acting on the 
constant-curvature slice~$\Sigma_\kappa$ parametrized by 
$(\theta,\phi)$, together with the translation generator 
$\hat{\mathcal{K}}_\varsigma=\partial_\varsigma$ of~$E_1$. Their cotangent lifts 
define a Hamiltonian action on $T^*(dS_3\times\mathbb{R})$ with momentum 
maps $\hat{\mathcal{J}}^{(\kappa)}_A = \hat p_\mu\,\hat{K}^{(\kappa)\mu}_A$, 
Eqs.~\eqref{eq:mommaps-ds}, and 
$\hat{\mathcal{J}}_\varsigma = \hat p_\varsigma$. These satisfy the Poisson 
bracket algebra
\begin{equation}
  \bigl\{\hat{\mathcal{J}}^{(\kappa)}_A,\,
  \hat{\mathcal{J}}^{(\kappa)}_B\bigr\} 
  = f_{AB}^{\;C}\,\hat{\mathcal{J}}^{(\kappa)}_C\,,
  \qquad
  \bigl\{\hat{\mathcal{J}}^{(\kappa)}_A,\,
  \hat{\mathcal{J}}_\varsigma\bigr\} = 0\,,
\end{equation}
cf.\ Eq.~\eqref{eq:LiealgJdsfol}. The geodesic Hamiltonian 
$\mathcal{H}=\tfrac{1}{2}\hat{g}^{\mu\nu}\hat p_\mu \hat p_\nu$ is 
$G$-invariant, so 
$\{\mathcal{H},\hat{\mathcal{J}}^{(\kappa)}_A\}=0$ and 
$\{\mathcal{H},\hat{\mathcal{J}}_\varsigma\}=0$, as required.

\medskip\noindent
\textit{Step~2: Identification of Casimirs.}
The Lie algebra $\mathfrak{h}_\kappa \in \{\mathfrak{so}(3),\, \mathfrak{iso}(2),\, \mathfrak{so}(2,1)\}$ has dimension~3 and its structure 
matrix $F_{ab}$ has generic rank~2, yielding one independent Casimir. As 
constructed in Sect.~\ref{subsec:canlift}, this is the quadratic 
invariant
\begin{equation}
  \hat{\mathcal{C}}_\kappa 
  = \bigl(\hat{\mathcal{J}}^{(\kappa)}_1\bigr)^{\!2}
  + \bigl(\hat{\mathcal{J}}^{(\kappa)}_2\bigr)^{\!2}
  + \kappa\bigl(\hat{\mathcal{J}}^{(\kappa)}_3\bigr)^{\!2}
  = \hat{p}^{\,2}_{\Sigma_\kappa}\,,
\end{equation}
Eq.~\eqref{eq:Cas-fol}. For the Abelian factor~$E_1$, the 
globally invariant Casimir under the full 
$E_1 = T(1)\rtimes O(1)$ action (including reflections 
$\varsigma\to-\varsigma$) is
\begin{equation}
  \hat{\mathcal{C}}_\varsigma 
  = \bigl(\hat{\mathcal{J}}_\varsigma\bigr)^{2} 
  = \hat p_\varsigma^{\,2}\,,
\end{equation}
Eq.~\eqref{eq:cas-momlong}. These two Casimirs, together with~$\rho$, 
are candidates for the complete set of independent variables.

\medskip\noindent
\textit{Step~3: Verification of cohomogeneity zero.}
This is the essential step. We must show that, after fixing 
$\hat{\mathcal{C}}_\kappa$ and $\hat{\mathcal{C}}_\varsigma$, the 
lifted $G$-action has cohomogeneity zero---i.e., acts transitively on 
each connected component of the resulting level set.

Consider the action restricted to the slice variables 
$(\theta,\phi,p_\theta,p_\phi)$. The four-dimensional cotangent bundle 
$T^*\Sigma_\kappa$ carries the lifted action of the three-dimensional 
group~$H_\kappa$. Since $H_\kappa$ acts transitively on the 
two-dimensional slice~$\Sigma_\kappa$---this is the defining property of 
a maximally symmetric space: $SO(3)$ acts transitively on~$\mathbb{S}^2$, 
$ISO(2)$ on~$\mathbb{R}^2$, and $SO(2,1)$ on~$\mathbb{H}^2$---the 
cotangent-lifted orbits in $T^*\Sigma_\kappa$ have dimension~3 at 
generic points (the full dimension of~$H_\kappa$), since the stabilizer 
of a generic point in $T^*\Sigma_\kappa$ is discrete. Hence these orbits 
have codimension $4-3=1$ in $T^*\Sigma_\kappa$, meaning there is 
exactly one continuous $H_\kappa$-invariant function of the slice 
phase-space variables. This invariant is the quadratic Casimir 
$\hat{\mathcal{C}}_\kappa = \hat{p}^{\,2}_{\Sigma_\kappa}$.

Now fix $\hat{\mathcal{C}}_\kappa = c_\kappa$ (a constant). The level 
set $\{\hat{p}^{\,2}_{\Sigma_\kappa}=c_\kappa\}\subset 
T^*\Sigma_\kappa$ is three-dimensional. Since $H_\kappa$ has 
dimension~3 and acts on this level set (because 
$\hat{\mathcal{C}}_\kappa$ is $H_\kappa$-invariant), and the orbits are 
generically three-dimensional, the action is transitive on each 
connected component of this level set. That is,
\begin{equation}
  \mathrm{cohom}\bigl(H_\kappa\curvearrowright 
  \{\hat{\mathcal{C}}_\kappa = c_\kappa\}\bigr) = 0\,.
\end{equation}

Independently, fixing 
$\hat{\mathcal{C}}_\varsigma = p_\varsigma^{\,2} = c_\varsigma$ removes 
the continuous freedom in~$\varsigma$ (invariance under 
$\partial_\varsigma$ already eliminates the coordinate~$\varsigma$ 
itself, and fixing the Casimir fixes $|p_\varsigma|$, leaving at most a 
discrete sign ambiguity).

Combining both sectors: after fixing 
$(\hat{\mathcal{C}}_\kappa,\hat{\mathcal{C}}_\varsigma)$, the quotient 
of the mass shell by $H_\kappa\times E_1$ is discrete on each connected 
component. The cohomogeneity-zero hypothesis of 
Proposition~\ref{prop:cohom0} is therefore satisfied.

\medskip\noindent
\textit{Step~4: $\rho$~is invariant---application of 
Proposition~\ref{prop:cohom0}.}
It remains to show that $\rho$ survives as an unconstrained parameter. 
The generators $\hat{\mathcal{K}}^{(\kappa)}_A$ are tangent to the constant-$\rho$ 
slices by construction: they act on $(\theta,\phi)$ only, so 
$\hat{K}^{(\kappa)}_A(\rho)=0$. Their cotangent lifts therefore satisfy 
$X_A[\rho]=0$, where $X_A$ is the Hamiltonian vector field of 
$\hat{\mathcal{J}}^{(\kappa)}_A$. Likewise, 
$\partial_\varsigma(\rho)=0$. Hence $\rho$ is constant along every 
$G$-orbit and is not constrained by the symmetry conditions 
$\{f,\hat{\mathcal{J}}^{(\kappa)}_A\}=0$ and $\{f,\hat{\mathcal{J}}_\varsigma\}=0$. It 
may therefore appear freely as an argument of~$f$.

\medskip\noindent
Applying Proposition~\ref{prop:cohom0} with Steps~1--3 verified, and 
adjoining the unconstrained parameter~$\rho$ from Step~4, we conclude 
that any smooth distribution function satisfying the symmetry conditions 
can be written locally as
\begin{equation}
  f = F\!\bigl(\rho,\;\hat{\mathcal{C}}_\kappa,\;
  \hat{\mathcal{C}}_\varsigma\bigr)
  = F\!\bigl(\rho,\;\hat{p}^{\,2}_{\Sigma_\kappa},\;
  p_\varsigma^{\,2}\bigr)\,,
\end{equation}
which is Eq.~\eqref{eq:reduced-form-dS}.\hfill$\square$

\section{Killing vectors of a 2d constant curvature manifold}
\label{app:diffgeom}

In this section, we outline some relevant results used in Sec.~\ref{sec:cotangent} which can be derived from standard methods of differential geometry~\cite{do1992riemannian,lee2018introduction}. The maximally symmetric submanifold $\Sigma_\kappa\subset\,$\ds with constant spatial curvature $\kappa$ is parameterized by coordinates $(\theta,\phi)$. Its induced metric is   
\begin{equation}
    \label{eq:indmet}
    \begin{split}
    d\Sigma_\kappa^2 &= h_{ij}dx^idx^j= d\theta^2+Q_\kappa^2(\theta)d\phi^2\,.
    \end{split}
\end{equation}
Equation~\eqref{eq:indmet} describes the canonical geometry of a two-dimensional space form of curvature $\kappa$, corresponding, respectively, to a sphere $\mathbb{S}^2$ ($\kappa=1$), an euclidian plane $\mathbb{R}^2$ ($\kappa=0$), or a hyperbolic plane  $\mathbb{H}^2$ ($\kappa=-1$), embedded in $dS_3$. For the induced metric~\eqref{eq:indmet} the non-zero Christoffel symbols are 
\begin{equation}
    \label{eq:Chris_indmet}
    \Gamma^\theta_{\phi\phi} =-Q_\kappa(\theta)Q'_\kappa(\theta)\,,\quad\quad\Gamma^\phi_{\phi\theta}=\Gamma^\phi_{\theta\phi} = \frac{Q'_\kappa(\theta)}{Q_\kappa(\theta)}\equiv G_\kappa(\theta)\,.
\end{equation}
In a two dimensional manifold, the Ricci scalar and the Gaussian curvatures are related by simply $\mathcal{R}=2\kappa$~\cite{do1992riemannian,lee2018introduction}. For the metric~\eqref{eq:indmet} this relation implies the following differential equation 
\begin{equation}
    Q''_\kappa + \kappa\, Q_\kappa(\theta) = 0, \quad\Rightarrow\quad 
\left(Q_\kappa' \right)^2 + \kappa\, Q^2_\kappa = 1,
\label{eq:gauss-ricci}
\end{equation}
Notice that, by solving the previous differential equation for $\kappa =\{1,0,-1\}$, corresponding to spherical, flat, and hyperbolic geometries respectively, and using as initial conditions $\{Q(0)=0,Q'_\kappa=1\}$, one obtains the three functions $Q_\kappa$ listed in Table~\ref{table:ds3defs}. For the case $\kappa=1$, the resulting induced metric~\eqref{eq:indmet} is maximally symmetric under \(\mathfrak{so}(3)\).  Consequently, one can anticipate that the induced metric $h_{ij}$ in Eq.~\eqref{eq:indmet} admits a maximally symmetric three-dimensional isometry algebra, corresponding to three independent Killing vectors.

On the two-dimensional submanifold $\Sigma_\kappa\subset\,$\ds with local coordinates $(\theta,\phi)$, any tangent vector field can be expressed in components as 
\begin{equation}
    \label{eq:Kill_ansatz}
    \mck_{A} = \sum_{i=1}^2\,\hat{\mathcal{K}}_A^{i,(\kappa)}e_i= \hat{\mathcal{K}}^{\theta,(\kappa)}_{A}(\theta,\phi)\,e_\theta + \hat{\mathcal{K}}^{\phi,(\kappa)}_{A}(\theta,\phi)e_\phi\,\quad\quad\quad\text{with}\,\,(e_\theta,e_\eta)=(\partial_\theta\,,\partial_\phi)\,,
\end{equation}
where the coordinate vectors $(\partial_\theta,\partial_\phi)$ form a local basis in the tangent space $\mathcal{T}_p\Sigma_\kappa$ at each point $p$. The Killing equations~\cite{do1992riemannian,lee2018introduction}, \(\nabla_i \mathcal{K}_j + \nabla_j \mathcal{K}_i = 0\), give the following set of coupled PDEs 
\begin{subequations}
\label{eq:Killeqs}
    \begin{align}
    \label{eq:kill1}
        &(\theta\theta):\quad \partial_\theta \mck _{A,\theta}=0\,,\quad\quad\Rightarrow\,\mck _{A,\theta}=\mck _{A,\theta}(\phi),\\
        \label{eq:kill2}
        &(\theta\phi):\quad\partial_\phi \mck _{A,\theta}+\partial_\theta(Q^2_\kappa(\theta)\, \mck _{A,\phi})-2G_\kappa(\theta)\, Q^2_\kappa(\theta)\, \mck _{A,\phi}=0\,,\,,\\
        \label{eq:kill3}
        &(\phi\phi):\quad\partial_\phi(Q^2_\kappa(\theta)\, \mck _{A,\phi})+G_\kappa(\theta)\, Q^2_\kappa(\theta)\mck _{A,\theta}=0\,.
    \end{align}
\end{subequations}
Since $0\leq\phi\leq 2\pi$ (see Table~\ref{table:ds3defs}), the most general solution\footnote{In principle, the most general solution would also include higher order harmonics as well, such as $\alpha_n \cos(n \phi) + \beta_n \sin(n \phi)$ with $n>1$.  However, one finds in this case that the two remaining equations \eqref{eq:kill2}-\eqref{eq:kill3} cannot be solved simultaneously when these higher order terms are included.} of Eq.~\eqref{eq:kill1} is
\begin{equation}
    \label{eq:kthsolgen}
    \mck _{A,\theta}(\phi) = \alpha\cos(\phi)+\beta\sin(\phi)\,+\gamma\,.
\end{equation}
where $\alpha$, $\beta$ and $\gamma$ are constants to be determined. It follows from this expression that, to satisfy Eq.~\eqref{eq:kill3}, $\mck _{A,\phi}$ must take the following generic functional form
\begin{equation}
    \label{eq:kphisolgen}
    \mck _{A,\phi}(\theta,\phi) = u(\theta)\,\sin\phi \,+\,v(\theta)\,\cos\phi\,+\,b_0(\theta)\,.
\end{equation}
Inserting this into Eq.~\eqref{eq:kill3} yields
\begin{equation}
    \label{eq:u-v-sols}
    u(\theta) =-G_\kappa(\theta)\alpha \,,\quad\quad v(\theta) = G_\kappa(\theta)\beta \,,
\end{equation}
so that
\begin{equation}
    \label{eq:kphisolgen2}
    \mck _{A,\phi}(\theta,\phi) = -G_\kappa(\theta)\alpha\,\sin\phi \,+\, G_\kappa(\theta)\beta\,\cos\phi\,+\,b_0(\theta)\,.
\end{equation}
Substituting \eqref{eq:kphisolgen2} into \eqref{eq:kill2} and equating the coefficients
of \(\sin\phi\), \(\cos\phi\), and the constant term yields two independent conditions:
\begin{subequations}
\begin{align}
-1 - (Q_\kappa Q'_\kappa)' + 2G_\kappa Q_\kappa Q'_\kappa &= 0, 
\label{cond:1}\\[2pt]
(Q^2_\kappa b_0)' - 2G_\kappa Q^2_\kappa b_0 &= 0.
\label{cond:2}
\end{align}
\end{subequations}
Using \((Q_\kappa Q'_\kappa)'=(Q'_\kappa)^2+Q_\kappa Q''_\kappa \) and \(G_\kappa Q_\kappa Q'_\kappa=(Q'_\kappa)^2\),
the first condition~\eqref{cond:1} simplifies to
\begin{equation}
-1 + (Q'_\kappa)^2 - Q_\kappa Q''_\kappa = 0.
\label{eq:cond_curv}
\end{equation}
For a surface of constant curvature, the metric function \(Q_\kappa(\theta)\)
satisfies~\eqref{eq:gauss-ricci};
substituting these into \eqref{eq:cond_curv} gives
\(-1+(Q'_\kappa)^2-Q_\kappa Q''_\kappa=-1+(Q'_\kappa)^2+\kappa Q^2_\kappa=0\),
so condition \eqref{eq:cond_curv} holds identically.
Hence, no additional restriction arises on the constants \(\alpha,\beta\).
The second condition \eqref{cond:2} integrates directly to
\(Q^2_\kappa b_0'=0\Rightarrow b_0(\theta)=\mathrm{const}\).
Finally, the constant term \(\gamma\) in \(\mck _{A,\phi}\)~\eqref{eq:kthsolgen}
must vanish from \eqref{eq:kill3}.
Indeed, with \(\gamma\neq0\), the first term in the RHS of Eq.~\eqref{eq:kill3} is \(2\pi\)-periodic and its average over \(\phi\) vanishes,
while \(Q_\kappa Q'_\kappa\gamma\) is \(\phi\)-independent.
Therefore, consistency requires \(Q_\kappa Q'_\kappa\gamma=0\), implying \(\gamma=0\).

Summarizing, the components of the most general Killing vector field~\eqref{eq:Kill_ansatz} for the induced metric~\eqref{eq:indmet} are
\begin{align}
\mck _{A,\theta}(\phi)
   &= \alpha\cos\phi + \beta\sin\phi,
   \\
\mck _{A,\phi}(\theta,\phi)
   &= -\,G_\kappa(\theta)\,\alpha\sin\phi
      + G_\kappa(\theta)\,\beta\cos\phi
      + b_0,
\end{align}
with constants \(\alpha,\beta,b_0\).
Choosing the triplets
\((\alpha,\beta,b_0)=(1,0,0),(0,1,0),(0,0,1)\)
gives the canonical basis
\begin{equation}
\label{eq:Killvects}
\begin{split}
\mck _1(\theta,\phi,\kappa)&=\cos\phi\,\partial_\theta-G_\kappa(\theta)\sin\phi\,\partial_\phi\,,\\
\mck _2(\theta,\phi,\kappa)&=\sin\phi\,\partial_\theta+G_\kappa(\theta)\cos\phi\,\partial_\phi,\\
\mck _3(\theta,\phi,\kappa)&=\partial_\phi\,.
\end{split}
\end{equation}
\subsection{Lie algebra}
\label{app:Lie}
Given the three Killing vectors~\eqref{eq:Killvects}, we determine their algebraic structure by explicitly evaluating the Lie bracket between different pairs of Killing vectors as follows
\begin{itemize}
    \item $\left[\mck _3,\mck _1\right]$: 
\begin{equation}
    \label{eq:Liealg1}
    \begin{split}
    \left[\mck _3,\mck _1\right]&=\partial_\phi(\cos\phi)\,\partial_\theta \,-\,G_\kappa(\theta)\partial_\phi(\sin\phi)\partial_\phi \,,\\
    &=-\left(\sin\phi\,\partial_\theta+G_\kappa(\theta)\cos\phi\,\partial_\phi,\right)\equiv -\mck _2\,.
    \end{split}
    \end{equation}
    \item $\left[\mck _3,\mck _2\right]$:
    \begin{equation}
        \label{eq:Liealg2}
        \begin{split}
        \left[\mck _3,\mck _1\right]&= \partial_\phi(\sin\phi)\,\partial_\theta \,+\,G_\kappa(\theta)\partial_\phi(\cos\phi)\partial_\phi \,,\\
        &=\left(\cos\phi\,\partial_\theta-G_\kappa(\theta)\sin\phi\,\partial_\phi,\right)\equiv \mck _1\,.
        \end{split}
    \end{equation}
    \item $\left[\mck _1,\mck _2\right]$: 
    \begin{equation}
        \label{eq:Liealg3}
        \begin{split}
        \left[\mck _1,\mck _2\right]&= \underbrace{\left(\mck _1(\sin\phi)-\mck _2(\cos\phi)\right)}_{\equiv 0}\partial_\theta \\
        &+ \left(\mck _1(G_\kappa(\theta)\cos\phi)+\mck _2(G_\kappa(\theta)\sin\phi)\right)\partial_\phi\,,\\
        &=\left(G'_\kappa(\theta) + G_\kappa^2(\theta) \right)\partial_\phi\,,\\
        &=-\kappa\,\partial_\phi \equiv -\kappa\,\mck _3\,,
        \end{split}
    \end{equation}
    where in the third line we used~\eqref{eq:gauss-ricci}.
\end{itemize}
In summary, the Killing vectors of a 2d constant curvature surface satisfy the following non-abelian Lie algebra
\begin{equation}
[\mck_1,\mck_2]=-\kappa\,\mck_3,\qquad
[\mck_2,\mck_3]=-\mck_1,\qquad
[\mck_3,\mck_1]=-\mck_2,
\label{eq:Liealg}
\end{equation}
The structure constants can be read directly from the commutation relations, yielding
\begin{equation}
    \label{eq:strct}
    \{f_{12}{}^{3},f_{23}{}^{1},f_{31}{}^{2}\}=\{-\kappa,-1,-1\}\,.
\end{equation}
Depending on the curvature parameter 
$\kappa$, the non-abelian Lie algebra~\eqref{eq:Liealg} corresponds respectively to 
\(\mathfrak{so}(3)\), \(\mathfrak{iso}(2)\) and \(\mathfrak{so}(2,1)\) algebras for $\kappa=\{1,0,-1\}$. Notably, one of the structure constants depends explicitly on the spatial curvature $\kappa$ of the 2d submanifold $\Sigma_\kappa$ emphasizing the deep correspondence between geometry and its associated algebraic structure. This dependence exemplifies how curvature encodes itself in the commutation relations of the symmetry generators.

\section{Anisotropic integrals}
\label{app:integrals}
The computation of the energy–momentum tensor for the exact solution of the Boltzmann equation, Eq.~\eqref{eq:boltzman_dssol}, requires evaluating several angular momentum integrals that appear naturally within the framework of anisotropic hydrodynamics (cf. Ref.~\cite{Martinez:2017ibh}). For completeness, we collect here the key integrals used throughout this work. The anisotropic integrals $\hat{I}_{nlq}$ are defined as
\begin{equation}
\label{eq:I_nql}
    \hat{I}_{nlq}=
   \int_{\pp(\rho)} \left(\,\hat{p}^\rho(\rho)\,\right)^{n}\,\left(\hat{p}_\varsigma\right)^l\, \left(\frac{\hat{p}_{\Sigma_\kappa}^2}{S_\kappa^2(\rho)}\right)^q\,f_{eq.}^{(\kappa)}\left(\frac{\hat{p}^\rho(\rho_0)}{\hat{T}_\kappa(\rho_0)}\right)\,,
\end{equation}
with $\hat{T}_{0,\kappa}\equiv \hat{T}_\kappa(\rho_0)$ and we are considering a generic equilibrium distribution function $f_{eq.}(x)=e^{-x}$. In addition, the $\rho$-dependent measure is $\int_{\pp(\rho)}\equiv\int d^3\pp/\left[(2\pi)^3\,S_\kappa(\rho)Q_\kappa(\theta)\,\pp^{\rho}\right]$. Upon the following change of variables
 \begin{subequations}
     \label{eq:change}
     \begin{align}
         \frac{\hat{p}_\theta}{S_\kappa(\rho_0)}&=\hat{T}_{0,\kappa}\,\lambda\,\sin\alpha\,\cos\beta\,,\\
         \frac{\hat{p}_\phi}{S_\kappa(\rho_0)\,Q_\kappa(\theta)}&=\hat{T}_{0,\kappa}\,\lambda\,\sin\alpha\,\sin\beta\,,\\
         \hat{p}_\varsigma&=\hat{T}_{0,\kappa}\,\lambda\,\cos\alpha\,,
     \end{align}
 \end{subequations}
the integral~\eqref{eq:I_nql} factorizes as follows
\begin{equation}
\label{eq:fact}
\hat{I}_{nlq}(\hat{T}_{0,\kappa};\rho,\rho_0)=\hat{J}_{nlq}(\hat{T}_{0,\kappa})\,\hat{R}_{nlq}\left(\,\left[\frac{S_\kappa(\rho_0)}{S_\kappa(\rho)}\right]^2\,\right)\,,
\end{equation}
where the functions $\hat{J}_{nlq}$ and $\hat{R}_{nlq}$ are given by
\begin{subequations}
\label{eq:J-R_functions}
\begin{align}
\label{eq:J-func}
\hat{J}_{nlq}(\hat{T}_{0,\kappa})&=\,\frac{\Gamma(n+l+2q+2)}{2\pi^2}\,\hat{T}_{0,\kappa}^{n+l+2q+2}\,,
\\
\label{eq:R-func}
\hat{R}_{nlq}\left(\,y\,\right)&=\,\frac{y^2}{2}\,\int_{-1}^1\,dx\,\left[y^2(1-x^2)+x^2\right]^{(n-1)/2}\,x^l\,\left[y^2(1-x^2)\right]^{2q}\,.
\end{align}
\end{subequations}
The integrals needed in this work are $\hat{I}_{200} = \hat{J}_{200}(\hat{T}_{0,\kappa})\,\hat{R}_{200}\left(y\right)$ and $\hat{I}_{020}=\hat{J}_{020}(\hat{T}_{0,\kappa})\,\hat{R}_{020}\left(y\right)$ - being $y=S_\kappa(\rho_0)/S_\kappa(\rho)$ - where the $\hat{J}$ and $\hat{R}$ functions are 
\begin{subequations}
\label{eq:integrals}
    \begin{align}
        \label{eq:ene-int}
        \hat{J}_{200}(\hat{T}_{0,\kappa})=\hat{J}_{020}(\hat{T}_{0,\kappa})&=\frac{3}{\pi^2}\hat{T}_{0,\kappa}^4\,,
        \\
        \label{eq:shear-int1}
        \hat{R}_{200}\left(y\right)&=\frac{y^2}{2}\left(1+y^2\,\frac{\tanh^{-1}\left(\sqrt{1-y^2}\right)}{\sqrt{1-y^2}}\right)\,,\\
\label{eq:shear-int2}
\hat{R}_{020}\left(y\right)&=\frac{1}{2}\,y^2\,\left(
        \frac{1}{1-y^2}-y^2\frac{\tanh^{-1}\left(\sqrt{1-y^2}\right)}{(1-y^2)^{3/2}}\right)\,.
    \end{align}
\end{subequations}
\section{The approximate cold-limit solution}
\label{app:coldODE}

In this appendix, we solve the differential equation Eq.~\eqref{eq:coldeqn}:
\begin{equation}
c\left[\frac{d\bar \pi^{(\kappa)}}{d\rho}+\frac{4}{3}\left(\bar \pi^{(\kappa)}(\rho)\right)^2\,R_\kappa(\rho)\right]
= \frac{4}{3}\,R_\kappa(\rho)
+ \frac{10}{21}c\,R_\kappa(\rho)\,\bar\pi^{(\kappa)}(\rho), 
\label{eq:genODE-app}
\end{equation}
for an arbitrary known Hubble-like scale factor $R_\kappa(\rho) = (1/S_\kappa(\rho))(dS_\kappa/d\rho)$. Introducing the logarithmic scale variable

\begin{equation}
u(\rho)\equiv\int^\rho R_\kappa(\rho')\,d\rho'=\ln S_\kappa(\rho)+\text{const},
\label{eq:u-def-app}
\end{equation}
we have
\begin{equation}
\frac{d\hat \pi^{(\kappa)}}{d\rho}=R_\kappa(\rho)\,\frac{d\pi}{du}.
\end{equation}
Substituting this into Eq.~\eqref{eq:genODE-app} and dividing by $R_\kappa(\rho)\neq0$
yields a Riccati equation with constant coefficients,
\begin{equation}
\begin{split}
\frac{d\hat \pi^{(\kappa)}}{du}
&= -\,\frac{4}{3}\,\left(\hat \pi^{(\kappa)}\right)^2 + \frac{10}{21}\,\hat \pi^{(\kappa)} + \frac{4}{3c}\,,\\
&=-\,\frac{4}{3}\,(\hat \pi^{(\kappa)}-\bar \pi_+)(\hat \pi^{(\kappa)}-\bar \pi_-),
\label{eq:riccati-app}
\end{split}
\end{equation}
where $\bar \pi_\pm$ are the fixed points of the Ricatti differential equation~\eqref{eq:riccati-app}
\begin{equation}
\bar \pi_{\pm}
= \frac{5}{28}\pm \frac{1}{28}\sqrt{25+\frac{784}{c}}\,.
\label{eq:pipm-app}
\end{equation}
Integrating Eq.~\eqref{eq:riccati-app} gives
\begin{equation}
\frac{1}{\bar \pi_+-\bar \pi_-}
\ln\!\left(
\frac{\hat \pi^{(\kappa)}-\bar \pi_+}{\hat \pi^{(\kappa)}-\bar \pi_-}
\right)
= -\frac{4}{3}u +\bar \pi_0 ,
\end{equation}
with integration constant $\bar \pi_0$ determined by the initial condition $\hat \pi^{(\kappa)}(\rho_0)$.  Solving for $\hat \pi^{(\kappa)}(u)$ and using $u=\ln S(\rho)$,
we obtain the explicit solution for arbitrary $S_\kappa(\rho)$
\begin{equation}
\bar \pi_{\kappa}^\mathrm{cold}(\rho)=
\frac{\bar \pi_+ - \bar \pi_0\,\bar \pi_-\,S_\kappa(\rho)^{-\frac{4}{3}(\bar \pi_+-\bar \pi_-)}}
     {1 - \bar \pi_0\,S_\kappa(\rho)^{-\frac{4}{3}(\bar \pi_+-\bar \pi_-)}}\,.
\label{eq:final-solution-app}
\end{equation}
The analytic structure of the cold-limit solution~\eqref{eq:final-solution-app} is that of a fractional linear (M\"obius) map. 

In Ref.~\cite{Marrochio:2013wla}, the authors identified the cold-limit solution of the traditional Israel–Stewart equation for the effective shear in the case of Gubser flow. Equation~\eqref{eq:genODE-app} reproduces their setup upon neglecting the term proportional to 
$10/21$ and using $S_\kappa(\rho)=\cosh\rho$. Solving the analogous IS equation for an arbitrary foliation - again dropping the 
$10/21$ term and following the same procedure outlined above - one finds that the cold-limit IS solution for a general \ds slicing takes the form
\begin{equation}
    \label{eq:coldIS-sol}
    \bar \pi_\kappa^\mathrm{IS-cold} = \frac{1}{\sqrt{c}}\, \tanh\left(\frac{4}{3\sqrt{c}}\left[\ln S_\kappa(\rho)-\bar \pi_0\,c\right]\right)\,.
\end{equation}
Notice that the full second-order equation~\eqref{eq:genODE-app} cannot be reduced to the IS form by any deformation.

\bibliographystyle{JHEP}
\bibliography{foliatedBoltzmann}

@article{debbasch2009general1,
  title={General relativistic {Boltzmann} equation, {I}: Covariant treatment},
  author={Debbasch, Fabrice and van Leeuwen, WA},
  journal={Physica A: Statistical Mechanics and its Applications},
  volume={388},
  number={7},
  pages={1079--1104},
  year={2009},
  publisher={Elsevier}
}

@article{debbasch2009general2,
  title={General relativistic {Boltzmann} equation, {II}: Manifestly covariant treatment},
  author={Debbasch, Fabrice and van Leeuwen, WA},
  journal={Physica A: Statistical Mechanics and its Applications},
  volume={388},
  number={9},
  pages={1818--1834},
  year={2009},
  publisher={Elsevier}
}

@article{Grozdanov:2025cfx,
    author = "Grozdanov, Sa{\v{s}}o",
    title = "{A family of three maximally symmetric boost-invariant flows in relativistic hydrodynamics}",
    eprint = "2510.10769",
    archivePrefix = "arXiv",
    primaryClass = "hep-th",
    month = "10",
    year = "2025"
}

@article{Denicol:2014xca,
    author = "Denicol, Gabriel S. and Heinz, Ulrich W. and Martinez, Mauricio and Noronha, Jorge and Strickland, Michael",
    title = "{New Exact Solution of the Relativistic Boltzmann Equation and its Hydrodynamic Limit}",
    eprint = "1408.5646",
    archivePrefix = "arXiv",
    primaryClass = "hep-ph",
    doi = "10.1103/PhysRevLett.113.202301",
    journal = "Phys. Rev. Lett.",
    volume = "113",
    number = "20",
    pages = "202301",
    year = "2014"
}

@article{Denicol:2014tha,
    author = "Denicol, Gabriel S. and Heinz, Ulrich W. and Martinez, Mauricio and Noronha, Jorge and Strickland, Michael",
    title = "{Studying the validity of relativistic hydrodynamics with a new exact solution of the Boltzmann equation}",
    eprint = "1408.7048",
    archivePrefix = "arXiv",
    primaryClass = "hep-ph",
    doi = "10.1103/PhysRevD.90.125026",
    journal = "Phys. Rev. D",
    volume = "90",
    number = "12",
    pages = "125026",
    year = "2014"
}

@article{Bazow:2015dha,
    author = "Bazow, D. and Denicol, G. S. and Heinz, U. and Martinez, M. and Noronha, J.",
    title = "{Analytic solution of the Boltzmann equation in an expanding system}",
    eprint = "1507.07834",
    archivePrefix = "arXiv",
    primaryClass = "hep-ph",
    reportNumber = "INT-PUB-15-038",
    doi = "10.1103/PhysRevLett.116.022301",
    journal = "Phys. Rev. Lett.",
    volume = "116",
    number = "2",
    pages = "022301",
    year = "2016"
}

@article{Bazow:2016oky,
    author = "Bazow, D. and Denicol, G. S. and Heinz, U. and Martinez, M. and Noronha, J.",
    title = "{Nonlinear dynamics from the relativistic Boltzmann equation in the Friedmann-Lema{\^\i}tre-Robertson-Walker spacetime}",
    eprint = "1607.05245",
    archivePrefix = "arXiv",
    primaryClass = "hep-ph",
    doi = "10.1103/PhysRevD.94.125006",
    journal = "Phys. Rev. D",
    volume = "94",
    number = "12",
    pages = "125006",
    year = "2016"
}

@article{Wang:2025wyh,
    author = "Wang, Yi and Zhao, Xuan and Xu, Zhe and Hu, Jin",
    title = "{Analytical Solution and Lie Algebra of the Relativistic Boltzmann Equation}",
    eprint = "2511.08652",
    archivePrefix = "arXiv",
    primaryClass = "cond-mat.stat-mech",
    month = "11",
    year = "2025"
}

@article{Hatta:2015kia,
    author = "Hatta, Yoshitaka and Martinez, Mauricio and Xiao, Bo-Wen",
    title = "{Analytic solutions of the relativistic Boltzmann equation}",
    eprint = "1502.05894",
    archivePrefix = "arXiv",
    primaryClass = "hep-th",
    reportNumber = "YITP-15-11",
    doi = "10.1103/PhysRevD.91.085024",
    journal = "Phys. Rev. D",
    volume = "91",
    number = "8",
    pages = "085024",
    year = "2015"
}

@article{Noronha:2015jia,
    author = "Noronha, Jorge and Denicol, Gabriel S.",
    title = "{Perfect fluidity of a dissipative system: Analytical solution for the Boltzmann equation in $\mathrm{AdS}_{2}\otimes \mathrm{S}_{2}$}",
    eprint = "1502.05892",
    archivePrefix = "arXiv",
    primaryClass = "hep-ph",
    doi = "10.1103/PhysRevD.92.114032",
    journal = "Phys. Rev. D",
    volume = "92",
    number = "11",
    pages = "114032",
    year = "2015"
}

@article{Hu:2024utr,
    author = "Hu, Jin",
    title = "{Analytical solution of the nonlinear relativistic Boltzmann equation}",
    eprint = "2411.16448",
    archivePrefix = "arXiv",
    primaryClass = "hep-ph",
    doi = "10.1007/JHEP07(2025)066",
    journal = "JHEP",
    volume = "2025",
    number = "07",
    pages = "066",
    year = "2025"
}

@article{Chen:2023vrk,
    author = "Chen, Shile and Shi, Shuzhe",
    title = "{Exact solution of Boltzmann equation~in a longitudinal expanding system}",
    eprint = "2311.09575",
    archivePrefix = "arXiv",
    primaryClass = "nucl-th",
    doi = "10.1103/PhysRevC.109.L051901",
    journal = "Phys. Rev. C",
    volume = "109",
    number = "5",
    pages = "L051901",
    year = "2024"
}

@article{Baym:1984np,
    author = "Baym, G.",
    title = "{Thermal Equilibration in Ultrarelativistic Heavy Ion Collisions}",
    doi = "10.1016/0370-2693(84)91863-X",
    journal = "Phys. Lett. B",
    volume = "138",
    pages = "18--22",
    year = "1984"
}

@book{Cercignani,
	Author = {Cercignani, C. and Medeiros Kremer, G.},
	Publisher = {Birkh\"auser Verlag (Basel, Switzerland)},
	Title = {The relativistic Boltzmann Equation: Theory and Applications},
	Year = {2002}}

@book{DeGroot:1980dk,
	Author = {de Groot, S. R. and van Leewen, W. A. and van Weert, C. G.},
	Publisher = {Elsevier North-Holland},
	Title = {Relativistic Kinetic Theory: principles and applications},
	Year = {1980}}

@article{BGK,
	Author = {Bhatnagar, P.~L. and Gross, E.~P. and Krook, M.},
	Doi = {10.1103/PhysRev.94.511},
	Journal = {Physical Review},
	Pages = {511-525},
	Title = {{A Model for Collision Processes in Gases. I. Small Amplitude Processes in Charged and Neutral One-Component Systems}},
	Volume = {94},
	Year = {1954}}

@article{Anderson:1974,
	Author = {J.L. Anderson and H.R. Witting},
	Doi = {10.1016/0031-8914(74)90355-3},
	Journal = {Physica},
	Number = {3},
	Pages = {466 - 488},
	Title = {{A relativistic relaxation-time model for the Boltzmann equation}},
	Volume = {74},
	Year = {1974}}

@article{KW-1,
	Author = {Krook, Max and Wu, Tai Tsun},
	Doi = {10.1103/PhysRevLett.36.1107},
	Issue = {19},
	Journal = {Phys. Rev. Lett.},
	Month = {May},
	Numpages = {0},
	Pages = {1107--1109},
	Publisher = {American Physical Society},
	Title = {Formation of Maxwellian Tails},
	Volume = {36},
	Year = {1976}}

@article{KW-2,
	Author = {Krook, Max and Wu, Tai Tsun},
	Doi = {http://dx.doi.org/10.1063/1.861780},
	Journal = {Physics of Fluids},
	Number = {10},
	Pages = {1589-1595},
	Title = {Exact solutions of the Boltzmann equation},
	Volume = {20},
	Year = {1977}}

@article{bobylev,
	Author = {Bobylev, A. V.},
	Journal = {Sov. Phys. Dokl.},
	Number = {34},
	Title = {Some properties of Boltzmann's equation for Maxwell molecules},
	Volume = {5/6},
	Year = {1984}}

@article{Sarbach:2013uba,
    author = "Sarbach, Olivier and Zannias, Thomas",
    title = "{The geometry of the tangent bundle and the relativistic kinetic theory of gases}",
    eprint = "1309.2036",
    archivePrefix = "arXiv",
    primaryClass = "gr-qc",
    doi = "10.1088/0264-9381/31/8/085013",
    journal = "Class. Quant. Grav.",
    volume = "31",
    pages = "085013",
    year = "2014"
}

@article{Sarbach:2013fya,
    author = "Sarbach, Olivier and Zannias, Thomas",
    editor = "Ure{\~n}a-L{\'o}pez, Luis A. and Becerril-B{\'a}rcenas, Ricardo and Linares-Romero, Rom{\'a}n",
    title = "{Relativistic Kinetic Theory: An Introduction}",
    eprint = "1303.2899",
    archivePrefix = "arXiv",
    primaryClass = "gr-qc",
    doi = "10.1063/1.4817035",
    journal = "AIP Conf. Proc.",
    volume = "1548",
    number = "1",
    pages = "134--155",
    year = "2013"
}

@article{Sarbach:2013hna,
    author = "Sarbach, Olivier and Zannias, Thomas",
    editor = "Mac{\'\i}as, Alfredo and Maceda, Marco",
    title = "{Tangent bundle formulation of a charged gas}",
    eprint = "1311.3532",
    archivePrefix = "arXiv",
    primaryClass = "gr-qc",
    doi = "10.1063/1.4861955",
    journal = "AIP Conf. Proc.",
    volume = "1577",
    number = "1",
    pages = "192--207",
    year = "2015"
}

@article{Denicol:2011fa,
    author = "Denicol, Gabriel S. and Noronha, Jorge and Niemi, Harri and Rischke, Dirk H.",
    title = "{Origin of the Relaxation Time in Dissipative Fluid Dynamics}",
    eprint = "1102.4780",
    archivePrefix = "arXiv",
    primaryClass = "hep-th",
    doi = "10.1103/PhysRevD.83.074019",
    journal = "Phys. Rev. D",
    volume = "83",
    pages = "074019",
    year = "2011"
}

@article{Rioseco:2016jwc,
    author = "Rioseco, Paola and Sarbach, Olivier",
    title = "{Accretion of a relativistic, collisionless kinetic gas into a Schwarzschild black hole}",
    eprint = "1611.02389",
    archivePrefix = "arXiv",
    primaryClass = "gr-qc",
    doi = "10.1088/1361-6382/aa65fa",
    journal = "Class. Quant. Grav.",
    volume = "34",
    number = "9",
    pages = "095007",
    year = "2017"
}

@phdthesis{rioseco2019relativistic,
  title={Relativistic Kinetic Theory with Applications in Astrophysics},
  author={Rioseco, Paola},
  year={2019},
  school={PhD thesis, Universidad Michoacana de San Nicol{\'a}s de Hidalgo}
}

@book{lee2018introduction,
	title = {Introduction to {Riemannian} manifolds},
	publisher = {Springer},
	author = {Lee, John},
	year = {2018},
}

@book{do1992riemannian,
  title={{Riemannian Geometry}},
  author={Do Carmo, Manfredo P.},
  year={1992},
  publisher={Birkh{\"a}user},
  address={Boston},
}

@article{Acuna-Cardenas:2021nkj,
    author = "Acu{\~n}a-C{\'a}rdenas, Rub{\'e}n O. and Gabarrete, Carlos and Sarbach, Olivier",
    title = "{An introduction to the relativistic kinetic theory on curved spacetimes}",
    eprint = "2106.09235",
    archivePrefix = "arXiv",
    primaryClass = "gr-qc",
    doi = "10.1007/s10714-022-02908-5",
    journal = "Gen. Rel. Grav.",
    volume = "54",
    number = "3",
    pages = "23",
    year = "2022"
}

@article{Gubser:2010ze,
    author = "Gubser, Steven S.",
    title = "{Symmetry constraints on generalizations of Bjorken flow}",
    eprint = "1006.0006",
    archivePrefix = "arXiv",
    primaryClass = "hep-th",
    reportNumber = "PUPT-2340",
    doi = "10.1103/PhysRevD.82.085027",
    journal = "Phys. Rev. D",
    volume = "82",
    pages = "085027",
    year = "2010"
}

@article{Gubser:2010ui,
    author = "Gubser, Steven S. and Yarom, Amos",
    title = "{Conformal hydrodynamics in Minkowski and de Sitter spacetimes}",
    eprint = "1012.1314",
    archivePrefix = "arXiv",
    primaryClass = "hep-th",
    reportNumber = "PUPT-2358",
    doi = "10.1016/j.nuclphysb.2011.01.012",
    journal = "Nucl. Phys. B",
    volume = "846",
    pages = "469--511",
    year = "2011"
}

@article{beltrametti1966number,
  title={{On the number of Casimir operators associated with any Lie group}},
  author={Beltrametti, EG and Blasi, A},
  journal={Physics Letters},
  volume={20},
  number={1},
  pages={62--64},
  year={1966},
  publisher={Elsevier}
}

@book{humphreys2012introduction,
  title={Introduction to Lie algebras and representation theory},
  author={Humphreys, James E},
  volume={9},
  year={2012},
  publisher={Springer Science \& Business Media}
}

@article{Florkowski:2013lya,
    author = "Florkowski, Wojciech and Ryblewski, Radoslaw and Strickland, Michael",
    title = "{Testing viscous and anisotropic hydrodynamics in an exactly solvable case}",
    eprint = "1305.7234",
    archivePrefix = "arXiv",
    primaryClass = "nucl-th",
    doi = "10.1103/PhysRevC.88.024903",
    journal = "Phys. Rev. C",
    volume = "88",
    pages = "024903",
    year = "2013"
}

@article{Florkowski:2013lza,
    author = "Florkowski, Wojciech and Ryblewski, Radoslaw and Strickland, Michael",
    title = "{Anisotropic Hydrodynamics for Rapidly Expanding Systems}",
    eprint = "1304.0665",
    archivePrefix = "arXiv",
    primaryClass = "nucl-th",
    doi = "10.1016/j.nuclphysa.2013.08.004",
    journal = "Nucl. Phys. A",
    volume = "916",
    pages = "249--259",
    year = "2013"
}

@misc{MartPlumb,
  author       = {Martinez, Mauricio and Plumberg, Christopher},
  title  = {\textit{In preparation}},
  year         = {2025}
}

@article{Martinez:2017ibh,
    author = "Martinez, M. and McNelis, M. and Heinz, U.",
    title = "{Anisotropic fluid dynamics for Gubser flow}",
    eprint = "1703.10955",
    archivePrefix = "arXiv",
    primaryClass = "nucl-th",
    doi = "10.1103/PhysRevC.95.054907",
    journal = "Phys. Rev. C",
    volume = "95",
    number = "5",
    pages = "054907",
    year = "2017"
}

@article{Heinz:2015cda,
    author = "Heinz, Ulrich W. and Martinez, Mauricio",
    title = "{Investigating the domain of validity of the Gubser solution to the Boltzmann equation}",
    eprint = "1506.07500",
    archivePrefix = "arXiv",
    primaryClass = "hep-ph",
    doi = "10.1016/j.nuclphysa.2015.08.009",
    journal = "Nucl. Phys. A",
    volume = "943",
    pages = "26--38",
    year = "2015"
}

@article{Banerjee:1989by,
    author = "Banerjee, B. and Bhalerao, R. S. and Ravishankar, V.",
    title = "{Equilibration of the Quark - Gluon Plasma Produced in Relativistic Heavy Ion Collisions}",
    reportNumber = "TIFR/TH/89-2",
    doi = "10.1016/0370-2693(89)91041-1",
    journal = "Phys. Lett. B",
    volume = "224",
    pages = "16--20",
    year = "1989"
}

@article{Muller:1967zza,
    author = "Muller, Ingo",
    title = "{Zum Paradoxon der Warmeleitungstheorie}",
    doi = "10.1007/BF01326412",
    journal = "Z. Phys.",
    volume = "198",
    pages = "329--344",
    year = "1967"
}

@article{Israel:1979wp,
    author = "Israel, W. and Stewart, J. M.",
    title = "{Transient relativistic thermodynamics and kinetic theory}",
    doi = "10.1016/0003-4916(79)90130-1",
    journal = "Annals Phys.",
    volume = "118",
    pages = "341--372",
    year = "1979"
}

@article{Soloviev:2025uig,
    author = "Soloviev, Alexander",
    title = "{Close encounters with attractors of the third kind}",
    eprint = "2510.26599",
    archivePrefix = "arXiv",
    primaryClass = "hep-th",
    month = "10",
    year = "2025"
}

@article{Marrochio:2013wla,
    author = "Marrochio, Hugo and Noronha, Jorge and Denicol, Gabriel S. and Luzum, Matthew and Jeon, Sangyong and Gale, Charles",
    title = "{Solutions of Conformal Israel-Stewart Relativistic Viscous Fluid Dynamics}",
    eprint = "1307.6130",
    archivePrefix = "arXiv",
    primaryClass = "nucl-th",
    doi = "10.1103/PhysRevC.91.014903",
    journal = "Phys. Rev. C",
    volume = "91",
    number = "1",
    pages = "014903",
    year = "2015"
}

@article{Baier:2007ix,
    author = "Baier, Rudolf and Romatschke, Paul and Son, Dam Thanh and Starinets, Andrei O. and Stephanov, Mikhail A.",
    title = "{Relativistic viscous hydrodynamics, conformal invariance, and holography}",
    eprint = "0712.2451",
    archivePrefix = "arXiv",
    primaryClass = "hep-th",
    reportNumber = "BI-TP-2007-29, INT-PUB-07-45, SHEP-07-47",
    doi = "10.1088/1126-6708/2008/04/100",
    journal = "JHEP",
    volume = "04",
    pages = "100",
    year = "2008"
}

@article{Bjorken:1982qr,
    author = "Bjorken, J. D.",
    title = "{Highly Relativistic Nucleus-Nucleus Collisions: The Central Rapidity Region}",
    reportNumber = "FERMILAB-PUB-82-044-THY, FERMILAB-PUB-82-044-T",
    doi = "10.1103/PhysRevD.27.140",
    journal = "Phys. Rev. D",
    volume = "27",
    pages = "140--151",
    year = "1983"
}

@article{Carinena:1998pun,
    author = "Carinena, Jose F. and Ramos, Arturo",
    title = "{Integrability of Riccati equation from a group theoretical viewpoint}",
    eprint = "math-ph/9810005",
    archivePrefix = "arXiv",
    doi = "10.1142/S0217751X9900097X",
    journal = "Int. J. Mod. Phys. A",
    volume = "14",
    pages = "1935",
    year = "1999"
}

@article{Bemfica:2017wps,
    author = "Bemfica, F{\'a}bio S. and Disconzi, Marcelo M. and Noronha, Jorge",
    title = "{Causality and existence of solutions of relativistic viscous fluid dynamics with gravity}",
    eprint = "1708.06255",
    archivePrefix = "arXiv",
    primaryClass = "gr-qc",
    doi = "10.1103/PhysRevD.98.104064",
    journal = "Phys. Rev. D",
    volume = "98",
    number = "10",
    pages = "104064",
    year = "2018"
}

@book{kisil2012geometry,
  title={Geometry of M\"obius transformations: Elliptic, parabolic and hyperbolic actions of $SL_2(\mathbb{R})$},
  author={Kisil, Vladimir V},
  year={2012},
  publisher={World Scientific}
}

@article{Tindall:2016try,
    author = "Tindall, J. and Torres-Rincon, J. M. and Rose, J. B. and Petersen, H.",
    collaboration = "SMASH",
    title = "{Equilibration and freeze-out of an expanding gas in a transport approach in a Friedmann{\textendash}Robertson{\textendash}Walker metric}",
    eprint = "1612.06436",
    archivePrefix = "arXiv",
    primaryClass = "hep-ph",
    doi = "10.1016/j.physletb.2017.04.080",
    journal = "Phys. Lett. B",
    volume = "770",
    pages = "532--538",
    year = "2017"
}

@article{MarsdenWeinstein1974,
  author = {Marsden, J. E. and Weinstein, A.},
  title = {Reduction of symplectic manifolds with symmetry},
  journal = {Rep. Math. Phys.},
  volume = {5},
  pages = {121--130},
  year = {1974}
}

@book{MarsdenRatiuBook,
  author = {Marsden, J. E. and Ratiu, T. S.},
  title = {Introduction to Mechanics and Symmetry},
  publisher = {Springer},
  year = {1999}
}

@book{AbrahamMarsden,
  author = {Abraham, R. and Marsden, J. E.},
  title = {Foundations of Mechanics},
  publisher = {Benjamin/Cummings},
  year = {1978}
}

@book{OrtegaRatiu,
  author = {Ortega, J.-P. and Ratiu, T. S.},
  title = {Momentum Maps and Hamiltonian Reduction},
  publisher = {Birkhäuser},
  year = {2004}
}

@article{Meyer1973,
  author = {Meyer, K. R.},
  title = {Symmetries and integrals in mechanics},
  journal = {Dynamical Systems},
  volume = {5},
  pages = {259--272},
  year = {1973}
}

@article{diaz2017cohomogeneity,
  title={Cohomogeneity one actions on anti de Sitter spacetimes},
  author={Diaz-Ramos, JC and Kashani, SMB and Vanaei, MJ},
  journal={Results in Mathematics},
  volume={72},
  number={1},
  pages={515--536},
  year={2017},
  publisher={Springer}
}

@article{berndt2017cohomogeneity,
  title={Cohomogeneity one actions on Minkowski spaces},
  author={Berndt, J{\"u}rgen and D{\'\i}az-Ramos, Jos{\'e} Carlos and Vanaei, Mohammad Javad},
  journal={Monatshefte f{\"u}r Mathematik},
  volume={184},
  number={2},
  pages={185--200},
  year={2017},
  publisher={Springer}
}

@book{cushman2015global,
  title={Global Aspects of Classical Integrable Systems},
  author={Cushman, R.H. and Bates, L.M.},
  isbn={9783034809184},
  url={https://books.google.com/books?id=OVHMCQAAQBAJ},
  year={2015},
  publisher={Springer Basel}
}

\end{document}